\documentclass[%
 reprint,
 superscriptaddress,
 amsmath,amssymb,
 aps,
 prab,
 floatfix,
]{revtex4-2}

\usepackage{graphicx}
\usepackage{dcolumn}
\usepackage{bm}
\usepackage{hyperref}
\hypersetup{
    colorlinks=true,
    citecolor=blue,
    linkcolor=blue,
    urlcolor=blue
    }

\usepackage[euler]{textgreek}
\usepackage{makecell}
\usepackage{multirow}

\let\oldhash\#
\renewcommand{\#}{\texttt{\oldhash}} 

\begin{document}

\preprint{APS/123-QED}

\title{Design of the third-generation lead-based neutron spallation target for the neutron time-of-flight facility at CERN}

\author{R.~Esposito}
\email{raffaele.esposito@cern.ch}
\affiliation{European Laboratory for Particle Physics (CERN), 1211 Geneva 23, Switzerland}
\affiliation{Thermomechanical Metallurgy Laboratory, PX Group Chair, Ecole Polytechnique Fédérale de Lausanne (EPFL), 2002 Neuchâtel, Switzerland}
\author{M.~Calviani}
\email{marco.calviani@cern.ch}
\author{O.~Aberle}
\author{M.~Barbagallo}
\affiliation{European Laboratory for Particle Physics (CERN), 1211 Geneva 23, Switzerland}
\author{D.~\surname{Cano-Ott}}
\affiliation{Centro de Investigaciones Energéticas Medioambientales y Tecnológicas, Madrid, Spain}
\author{N.~Colonna}
\affiliation{Istituto Nazionale di Fisica Nucleare, Bari, Italy}
\author{T.~Coiffet}
\affiliation{European Laboratory for Particle Physics (CERN), 1211 Geneva 23, Switzerland}
\author{C.~\surname{Domingo-Pardo}}
\affiliation{Instituto de Física Corpuscular, CSIC-Universitat de Valencia, Spain}
\author{F.~Dragoni}
\author{R.~\surname{Franqueira~Ximenes}}
\author{L.~Giordanino}
\author{D.~Grenier}
\affiliation{European Laboratory for Particle Physics (CERN), 1211 Geneva 23, Switzerland}
\author{F.~Gunsing}
\affiliation{CEA Saclay, Irfu, Gif-sur-Yvette, France}
\author{K.~Kershaw}
\affiliation{European Laboratory for Particle Physics (CERN), 1211 Geneva 23, Switzerland}
\author{R.~Logé}
\affiliation{Thermomechanical Metallurgy Laboratory, PX Group Chair, Ecole Polytechnique Fédérale de Lausanne (EPFL), 2002 Neuchâtel, Switzerland}
\author{V.~Maire}
\author{P.~Moyret}
\author{A.~\surname{Perez~Fontenla}}
\author{A.~\surname{Perillo-Marcone}}
\author{F.~Pozzi}
\author{S.~Sgobba}
\author{M.~Timmins}
\author{V.~Vlachoudis}
\affiliation{European Laboratory for Particle Physics (CERN), 1211 Geneva 23, Switzerland}

\collaboration{for the n\_TOF Collaboration}


\begin{abstract}
The neutron time-of-flight (n\_TOF) facility at the European Laboratory for Particle Physics (CERN) is a pulsed white-spectrum neutron spallation source producing neutrons for two experimental areas: the Experimental Area~1 (EAR1), located 185~m horizontally from the target, and the Experimental Area~2 (EAR2), located 20~m above the target. The target, based on pure lead, is impacted by a high-intensity 20\nobreakdash-GeV/\textit{c} pulsed proton beam. The facility was conceived to study neutron-nucleus interactions for neutron kinetic energies between a few meV to several GeV, with applications of interest for nuclear astrophysics, nuclear technology, and medical research. After the second-generation target reached the end of its lifetime, the facility underwent a major upgrade during CERN's Long Shutdown 2 (LS2, 2019--2021), which included the installation of the new third-generation neutron target. The first and second-generation targets were based on water-cooled massive lead blocks and were designed focusing on EAR1, since EAR2 was built later. The new target is cooled by nitrogen gas to avoid erosion-corrosion and contamination of cooling water with radioactive lead spallation products. Moreover, the new design is optimized also for the vertical flight path and EAR2. This paper presents an overview of the target design focused on both physics and thermo-mechanical performance, and includes a description of the nitrogen cooling circuit and radiation protection studies.
\end{abstract}

\maketitle



\section{\label{sec:introduction}Introduction and motivations}

\begin{figure*}[t]
    \centering
    \includegraphics[width=\textwidth]{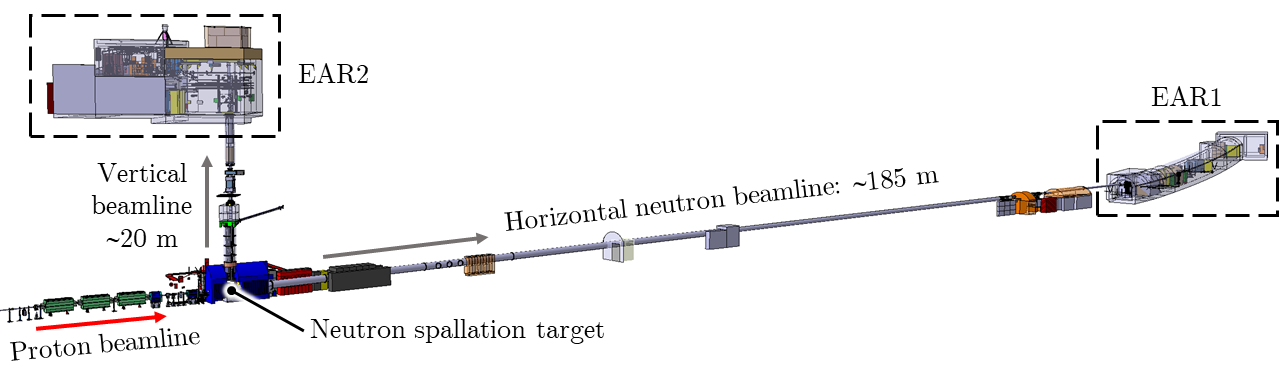}
    \caption{The n\_TOF Facility. A high-intensity proton beam from the CERN's Proton Synchrotron collides with a pure lead target, producing neutrons that travel along two flight paths toward two experimental areas: EAR1 and EAR2. EAR1 is located 185~m from the target, while EAR2 is located 20~m above the target.}
    \label{1-1_n_TOF}
\end{figure*}

The neutron time-of-flight (n\_TOF) facility at the European Laboratory for Particle Physics (CERN) is a neutron source capable of providing high-intensity pulsed white-spectrum neutrons covering almost eleven orders of magnitude, from thermal neutrons to several GeV. Neutrons are produced via spallation mechanism~\cite{Filges09} from the interaction between a pure-lead target and pulsed 20\nobreakdash-GeV/\textit{c} proton bunches from the Proton Synchrotron (PS) accelerator ring at CERN. The generated neutrons travel inside vacuum tubes along two flight paths directed to two experimental areas, Experimental Area~1 (EAR1) and Experimental Area~2 (EAR2), where cutting-edge experimental setups for neutron-induced reaction studies are in place. EAR1 is located at the end of a horizontal beamline, 185~m from the spallation target, while EAR2 is located at the end of a vertical beamline, 20~m above the target (Fig.~\ref{1-1_n_TOF}).
The facility was conceived to study neutron-nucleus interactions with applications of interest for nuclear astrophysics, nuclear technology, and medical research~\cite{Rubbia98,Mosconi10,Colonna15,Damone18,Guerrero20}.

The facility operated with EAR1 and a first-generation spallation target (Target~\#1) from 2000 to 2004~\cite{Borcea03}. A second-generation target (Target~\#2) operated from 2008 to the end of 2018. Even though the second target included many upgrades if compared with the first one, both targets were cooled by water in direct contact with pure lead~\cite{Esposito20}.

For the first target, the cooling water was used as moderator. An independent moderator assembly made from aluminum alloy (EN AW-5083 H111) was added to the second-generation target to separate the cooling water from the liquid used for moderation. The moderator liquid was then switched to borated water (1.28~wt\% with $^{10}$B-enriched boron at 99\%), which reduced the $\gamma$-ray background for the neutron capture measurements in EAR1~\cite{Barbagallo13} (the $\gamma$-ray background in n\_TOF is described in detail in Sec.~\ref{sec:physics:background}).

The second-generation target was designed considering an estimated lifetime of ten years. In the last years of target operation, an increase in the cooling water activity was detected, and preliminary signs of corrosion of the neutron window were observed by endoscopic inspections~\cite{Catherall15}. Moreover, during 2014, the second experimental area (EAR2) was built 20~m above the spallation target, although the neutron beam characteristics in EAR2 were not optimal since the target shape was not conceived for this experimental area.

All these aspects triggered the design of a new third-generation spallation target (Target~\#3), to be installed during the Long Shutdown~2 (LS2), a three-year stop of the CERN accelerators, and with the objective of starting operation in 2021. High-purity lead has been kept as core material owing to its superior performances in terms of reduced photon background with respect to other spallation materials, due to its very low neutron capture cross-sections.

Target~\#3 was designed with the assumption of a beam momentum of 20~GeV/\textit{c}, with a maximum number of protons per pulse equal to 10$^{13}$ (equivalent to a pulse kinetic energy of 32~kJ). The minimum pulse period is 1.2~s and the maximum average intensity allowed is 1.67$\times$10$^{12}$~p$^+$/s, corresponding to an average power on target of 5.4~kW. With a pulse duration of 7~ns (RMS), this yields a peak deposited power of 1.8~TW. The beam size on target is assumed to be 15~mm (RMS).

\subsection{\label{sec:introduction:neutron_sources}Pulsed white neutron sources}

Several white neutron sources are operating worldwide, each of them with specific characteristics and features, depending on the specific user requests~\cite{Colonna18}. However, only a few of them are pulsed and thus optimized for time-of-flight measurements. They are generally characterized by relatively high instantaneous neutron flux with very reduced photon background. The GELINA facility at the Joint Research Centre in Geel (Belgium) is based on a linear electron accelerator impinging on a mercury-cooled uranium target~\cite{Salome81, Febvre09}. The Gaerttner LINAC Center at the Rensselaer Polytechnic Institute in the United States employs a water-cooled tantalum target impacted by an electron beam~\cite{Danon95,Overberg99}. Tantalum-clad tungsten targets are employed in the Lujan Center at the Los Alamos Neutron Source Center (LANSCE) in the United States~\cite{Nowicki17}.

Neutron production targets based on lead are employed at the Paul Scherrer Institute (PSI) in Switzerland~\cite{Bauer10} and in the neutron time-of-flight setup nELBE at the Helmholtz-Zentrum Dresden-Rossendorf in Germany~\cite{Beyer13}. In the case of the SINQ source, the lead is solid without beam but operates in liquid state inside a Zircalloy cladding during operation~\cite{Thomsen11,Kiselev15,Wohlmuter20}, while in case of nELBE Pb is in liquid state at all times. Solid lead targets for neutron production are utilized for the Lead Slowing Down Spectrometer (LSDS) at the LANSCE accelerator~\cite{Rochman05}.


\section{\label{sec:physics}Physics reach}

The n\_TOF facility has been at the forefront of neutron physics since 2001~\cite{Gunsing16}. The design of Target~\#3 aims to further expand the measuring capabilities of the facility~\cite{Esposito20}. The combination of a high-energy and high-intensity proton beam, long flight paths, specifically designed optical elements along the beamlines, and an innovative spallation target results in high quality neutron beams making n\_TOF a unique facility. The main features of the facility are summarized in Table~\ref{T2-1_key_features}.

\begin{table}[h]
    \caption{\label{T2-1_key_features}Key features of the n\_TOF facility~\cite{Barbagallo13,Sabate17}.}
    \begin{tabular}{ r r l l }
    \hline\hline
    \multicolumn{2}{r}{\thead[r]{Quantity}} & \thead[l]{EAR1} & \thead[l]{EAR2} \\
    \hline
    \multicolumn{2}{r}{\makecell*[r]{Neutron flux\\(n/bunch)}} & 10$^6$ & 10$^8$ \\
    \multirow{2}{*}{\makecell*[r]{Energy range}} & min. & subthermal & subthermal \\
    & max. & 1~GeV & 100~MeV \\
    \multicolumn{2}{r}{\makecell*[r]{Best resolution\\(\textDelta E/E)}} & 10$^{-4}$ & 10$^{-3}$ \\
    \hline\hline
    \end{tabular}
\end{table}

Extensive Monte Carlo studies, performed by means of the \textsc{FLUKA} simulation package~\cite{Ferrari05,Bohlen14}, were conducted in parallel with the engineering design of Target~\#3 to help solving the issues of Target~\#2 mentioned in Sec.~\ref{sec:introduction} and further improve physics performance. A detailed and comprehensive \textsc{FLUKA} model of the target assembly was developed, based on the engineering design described in the next sections. The guiding goals of the Monte Carlo studies were, compared to Target~\#2: (1) Preserve intensity and shape of the neutron flux in the direction of EAR1. (2) Improve the resolution function for EAR2. (3) Reduce the $\gamma$-ray background in EAR1.

The next subsections provide an account of these optimization studies and their results.

\subsection{\label{sec:physics:neutron}Neutron fluence}

The key feature for both experimental areas is the high instantaneous neutron flux~\cite{Guerrero13,Barbagallo13,Sabate17}, of fundamental importance for measuring cross-sections of radioactive samples or samples available in small quantities. The integrated flux of 10$^6$ and 10$^8$ neutrons/bunch in EAR1 and EAR2, respectively, is due to the high intensity and high energy of the primary proton beam and the overall mass of the target (especially its lead component).

\begin{figure}[b]
    \centering
    \includegraphics[width=\columnwidth]{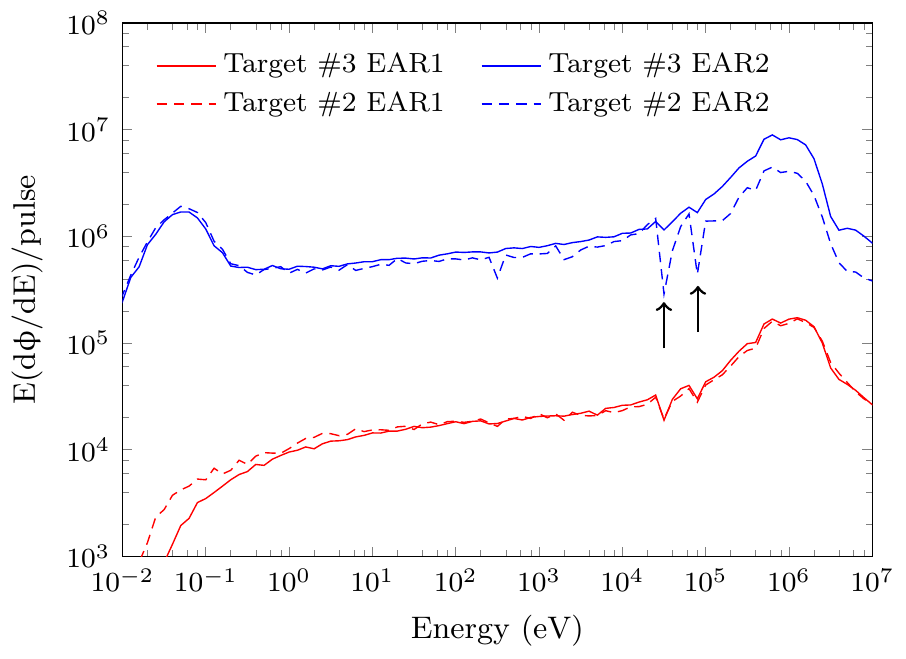}
    \caption{\textsc{FLUKA} simulation of the neutron flux in the two experimental areas for Target~\#2 and Target~\#3. The neutron flux in the energy range of interest is similar in EAR1 and much better for Target~\#3 in EAR2, especially considering the absence of absorption dips in the tens of keV range.}
    \label{2-1_neutron_flux_areas}
\end{figure}

As shown in Fig.~\ref{2-1_neutron_flux_areas}, Target~\#3 produces values of the neutron flux similar to Target~\#2.

The shape of the neutron beam and, to some extent, its intensity, also depend on the moderator systems. When compared to Target~\#2, Target~\#3 has a slightly thinner layer of moderator liquid (water or borated water) crossed by neutrons directed towards EAR1. This results in an additional contribution of non-moderated neutrons at the evaporation peak (hundreds keV to several MeV) and, consequently, in a slightly lower neutron flux in the thermal and epithermal regions ($\leq$\,100~eV). Overall, the neutron flux reaching EAR1 with Target~\#3 is 10\% lower than with Target~\#2. Regarding EAR2, Target~\#3 leaves the neutron flux in the range 25~meV--100~keV essentially unchanged, while a factor 2-3 higher is expected at the evaporation peak. The higher flux is only indirectly related to the new target design, as it depends mainly on the new design of the vacuum window coupling the target to the beamline.

Moreover, the smaller amount of aluminum present in the structure of the Target~\#3 ensures a substantial reduction of the related absorption dips in the flux, as indicated by the black arrows in Fig.~\ref{2-1_neutron_flux_areas}.

\subsection{\label{sec:physics:RF}Resolution function}

The neutron-induced reaction cross-sections under investigation at n\_TOF exhibit narrow resonances in the eV--keV region. The energy resolution of the facility is of utmost importance to accurately determine the cross-sections in this energy range. Energy resolution essentially depends on the flight path of the two beamlines and on the neutron moderation.
The resolution function can be expressed as the uncertainties d$\lambda$ in the equivalent flight path distance, and it is purely dictated by the geometry of the moderator. In the energy range of interest it is characterized by a dominant peak having a full width at half maximum (FWHM) of 2~cm, due to the pronounced slowdown in water, and a relatively small but long tail (up to several meters), due to the moderation in the lead core.

\begin{figure}[h]
    \centering 
    \includegraphics[width=\columnwidth]{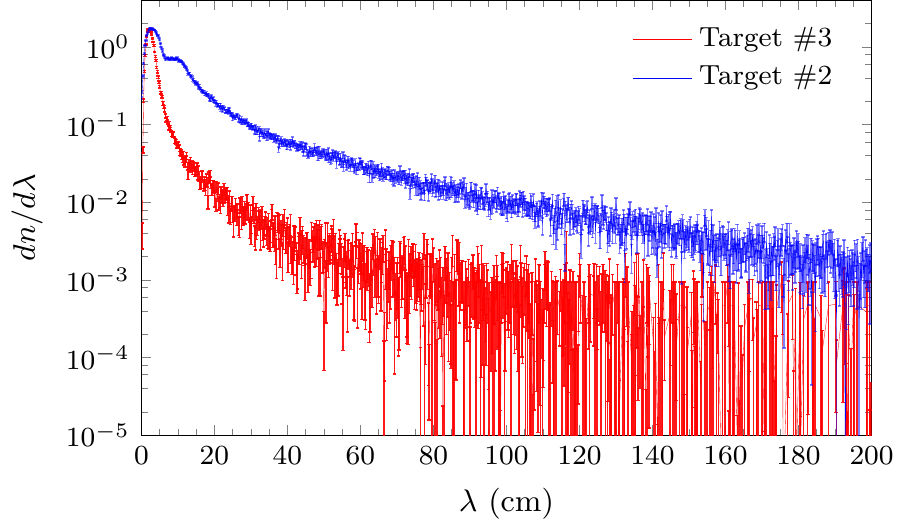}
    \caption{Distribution of the moderation length $\lambda$, in EAR2, for neutrons in the energy range 1--10 eV. Target~\#2 exhibits a double-peak structure leading to substantial limitations on the measurements in EAR2, whereas Target~\#3 exhibits a single peak with narrower FWHM, which is better suited for high-resolution measurements.}
    \label{2-2_resolution_function_EAR2_comparison}
\end{figure}

\textsc{FLUKA} simulations showed that the optimum, in terms of FWHM, can be reached with water layers of 4~cm for both EAR1 and EAR2 moderators. While the new design does not show any difference in the EAR1 resolution function when compared to Target~\#2~\cite{Guerrero13}, it significantly improves the scenario for EAR2.
In the Target~\#3 design, particular attention was paid in making the neutrons traveling towards EAR2 pass through a uniform layer of moderator water. This was not the case for Target~\#2, built before the construction of the second beamline and therefore not optimized for it. Figure~\ref{2-2_resolution_function_EAR2_comparison} shows the comparison of the $\lambda$ distribution of neutrons in EAR2 in the energy range 1--10~eV for the two targets, revealing a double-peak structure for Target~\#2 that led to substantial limitations on the measurements in EAR2. The double-peak structure is not present in the $\lambda$ distribution of Target~\#3, characterized by a much better FWHM around the peak.

\subsection{\label{sec:physics:background}Background}

A low background is a fundamental prerequisite to carry out high-quality cross-section measurements. At n\_TOF, the two main components of background related to the spallation target are neutrons and $\gamma$-rays.

The neutron contribution is proportional to the neutron flux and also depends on the optical elements placed in the downstream beamlines (i.e., collimators), so it is essentially unchanged in the new design. The contribution of $\gamma$-rays is strongly dependent on the target assembly, in particular on the moderator system and on the structural materials of the target. The $\gamma$-rays reaching the experimental areas can be divided into two contributions: \textit{prompt} and \textit{delayed}. The threshold between the two has been set at 900~ns for EAR1 and 200~ns for EAR2.

The prompt $\gamma$-rays reach EAR1 within 900~ns and EAR2 within 200ns. Their amount is directly proportional to the length of lead crossed by the primary proton beam, so no sizable differences are expected in the new target design with regard to EAR1. For EAR2, on the other hand, the coupling between target and beamline has been improved with a new optimized shape of the vacuum chamber, so a larger portion of the spallation target is seen by the experimental area and the $\gamma$-ray contribution to the background increases by a factor of 6. To compensate for the higher background, a 5\nobreakdash-cm thick lead plate has been integrated just above the target core and below the EAR2 moderator. \textsc{FLUKA} simulations indicate that the addition of this lead plate is enough to keep the prompt $\gamma$-ray contribution to the background in EAR2 as low as for Target~\#2, with no significant reduction in the neutron flux reaching the area.

\begin{figure}[h]
    \centering
    \includegraphics[width=\columnwidth]{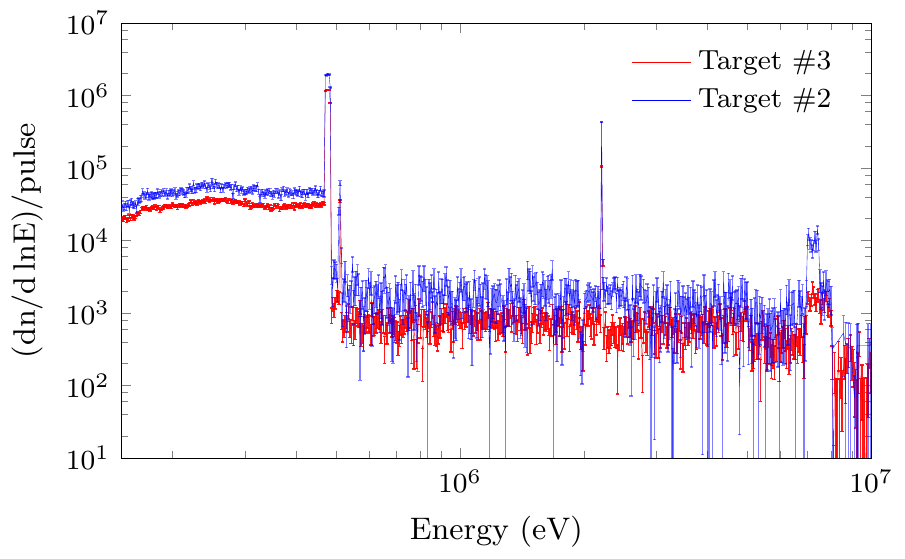}
    \caption{Delayed $\gamma$-ray distribution in EAR1 for Target~\#2 (blue) and Target~\#3 (red). A reduction by a factor 4-5 is visible in Target~\#3 at the 2.2-MeV energy peak and at the 7-MeV transition lines.}
    \label{2-3_background_EAR1_comparison}
\end{figure}

The delayed $\gamma$-rays reach EAR1 and EAR2 at a time-of-flight longer than 900~ns and 200~ns, respectively. They are generated from radiative capture reactions of neutrons in the hydrogen of the moderator water and in the structural materials of the target. The former can be inhibited using borated water (due to thermal neutron capture in $^{10}$B). For this reason, Target~ \#3 includes two decoupled moderator circuits, with the possibility of changing independently the circulating moderator liquid. In addition, considerable effort was put into minimizing the structural materials of the target that are sources of delayed $\gamma$-rays. Figure~\ref{2-3_background_EAR1_comparison} shows the \textsc{FLUKA} results of the delayed $\gamma$-ray background expected in EAR1 in comparison to Target~\#2, revealing a background lower by a factor 4-5, in particular at the 2.2\nobreakdash-MeV energy peak and at the 7\nobreakdash-MeV transition lines.

Regarding EAR2, $\gamma$-ray background arises mainly from the interaction of neutrons with the collimator at the entrance of the experimental area and with the beam dump in the roof of the building. Therefore, the design of the target does not impact the delayed $\gamma$-ray background in EAR2.

\section{\label{sec:design}Mechanical design}

\begin{figure*}[t]
    \centering
    \includegraphics[width=\textwidth]{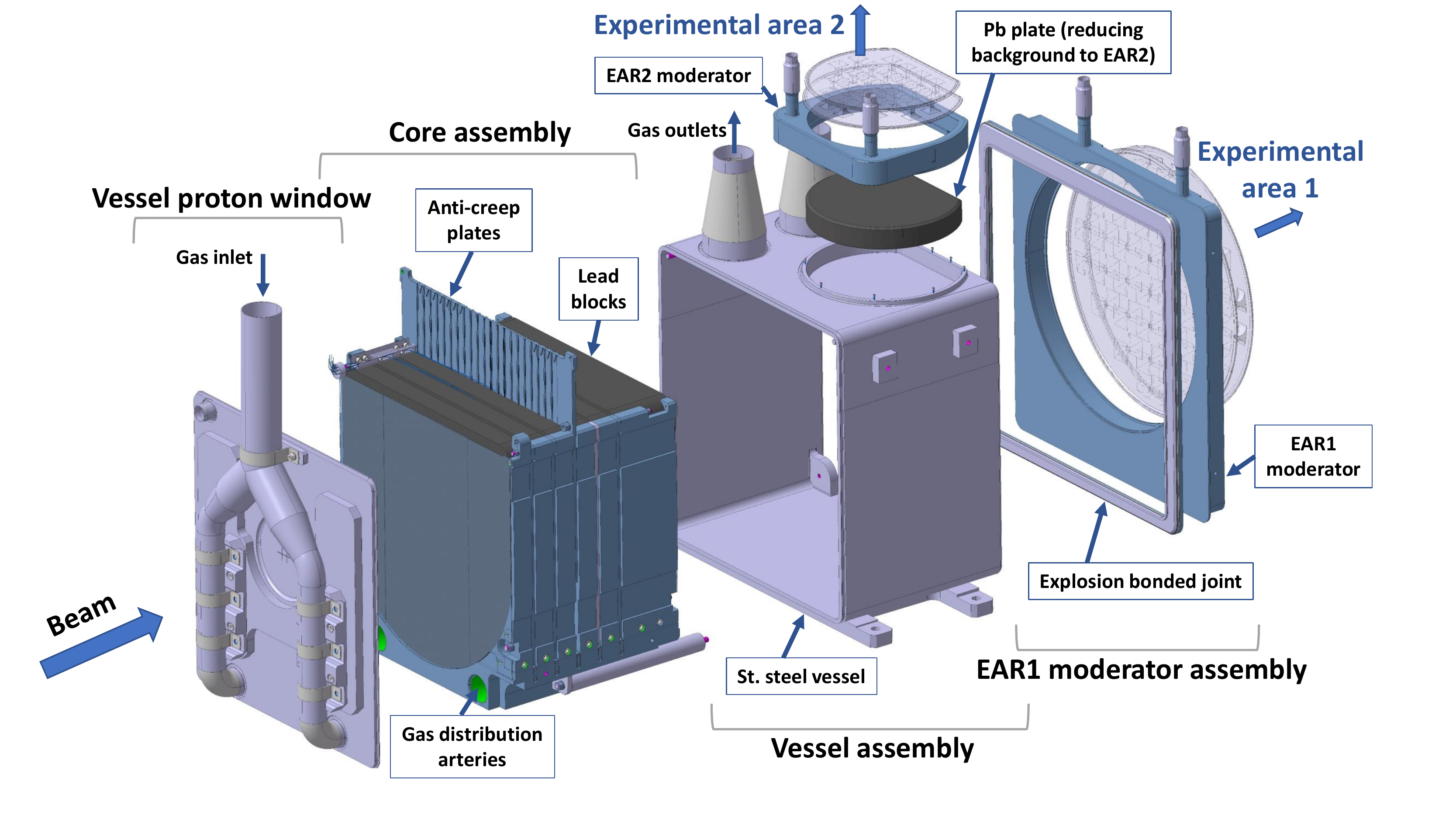}
    \caption{Exploded model of the target. The cooling nitrogen gas flows through the channels machined into the anti-creep plates. The inner core is enclosed in an AISI 316L stainless steel vessel. The two moderators are in aluminum EN AW-5083 H112. The EAR1 moderator is bonded to the stainless steel vessel by a bimetallic transition obtained by explosive bonding.}
    \label{3-1_target_assembly}
\end{figure*}

\begin{figure}[h]
    \centering
    \includegraphics[width=0.91\columnwidth]{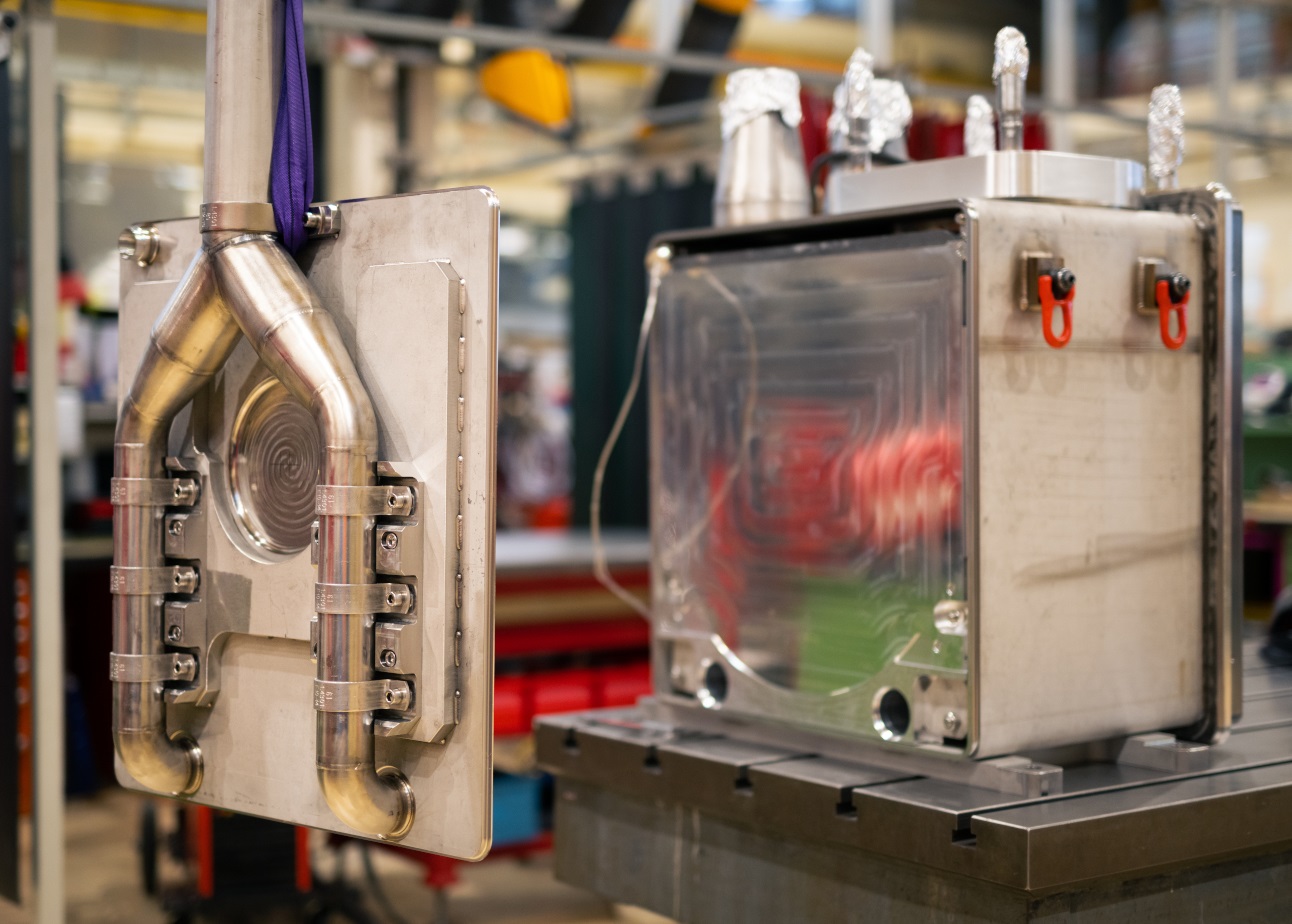}
    \caption{Photo of the n\_TOF Target~\#3 core inside the stainless steel vessel before closing the proton side of the vessel by TIG welding.}
    \label{3-2_photo_target}
\end{figure}

The new Target~\#3 assembly is shown, in exploded view, in Fig.~\ref{3-1_target_assembly}: housed inside a stainless steel vessel, six lead slices are cooled by gaseous nitrogen and are supported by precisely machined anti-creep plates. These are made from aluminum alloy and include the channels through which the cooling fluid flows. The cooling gas is distributed through two main arteries inside a cradle made from aluminum alloy, which supports the lead core from below. Connected to the vessel, two moderator containers, made from aluminum alloy, are positioned on the path of the neutrons directed to the two experimental areas. The bond between the stainless steel vessel and the aluminum moderator is obtained by an explosive-bonded joint. The lead slices are 5-cm thick, with the exception of the slice close to the EAR1 moderator, which is 15-cm thick. These values were selected to obtain a reasonable compromise between the total target length and lead surface exposed to the coolant. \textsc{FLUKA} simulations showed a conspicuous reduction in $\gamma$-ray background and an increase in neutron flux by merging the 15~cm of lead closest to the EAR1 moderator into a single slice. Figure~\ref{3-2_photo_target} shows a photo of the target core inside the stainless steel vessel before the proton side of the vessel was welded on to close it.

\subsection{\label{sec:design:core}Lead core and anti-creep structure}

The main target core is composed of six slices of high-purity lead supported by an aluminum (EN AW-6082 T6) structure that includes anti-creep plates between the slcies (Fig.~\ref{3-1_target_assembly}). The lead grade is UNS L50006, with a minimum purity of 99.98~wt\%. The anti-creep aluminum structure also plays a determinant role in the gas distribution thanks to channels machined to distribute optimally the nitrogen gas between the lead slices.

To ensure a correct assembly of the anti-creep plates between the lead slices, the gap between them must be within 45~\textmu m and 195~\textmu m. The minimum value takes the thermal expansion of the lead slices during operation into account, while the maximum value is required to avoid any negative impact on the nitrogen flow. This requirement entails the anti-creep plates to be manufactured with tight dimensional tolerances with respect to thickness (9.85\,$\pm$\,0.05~mm) and tight geometrical tolerances (flatness lower than 1~mm). The large size of the plates (0.6\,m\,$\times$\,0.6\,m) made this requirement particularly difficult to keep while milling the channels, since each milling step can bring distortions to the plate. The accumulation of deformations along the same direction was avoided by changing the side of the plate after each milling step. Furthermore, a dedicated vacuum table holding the plates during the milling procedure was manufactured, and the whole procedure was tuned and tested by first building a prototype of the plate.

\subsection{\label{sec:design:vessel_moderators}Vessel and Moderators design}

The target core is contained in a stainless steel (AISI 316L) vessel designed to hold a nominal nitrogen pressure of 0.5~bar. Made by first welding two u-shaped parts, the upstream side is closed by a plate equipped with the nitrogen inlet pipes, welded once the target core was inserted (Fig.~\ref{3-1_target_assembly}). A beam window was machined at the beam impact area by locally reducing the vessel thickness down to 3~mm to avoid significant heat dissipation due to the interaction between the proton beam and the vessel.
An additional stainless steel neutron window (4\nobreakdash-mm thick) was welded by electron beam to the top of the vessel and supports the lead plate that reduces the $\gamma$-ray background in EAR2 mentioned in Sec.~\ref{sec:physics:background}.
The two moderators are connected to two independent water circuits at a nominal pressure of 2.5~bar. They are made from forged aluminum EN AW-5083 H112. This alloy was chosen for its good mechanical properties, their stability after welding, good general corrosion resistance, and the ability to withstand the stresses induced by nitrogen pressure and thermal gradients.
The EAR2 moderator is located outside the vessel, placed on top of the 5\nobreakdash-cm lead plate above the target core. The moderator is bolted to the vessel and sealed at the top edge to prevent moisture building up between it and the lead plate.

\begin{figure}[h]
    \centering
    \includegraphics[width=\columnwidth]{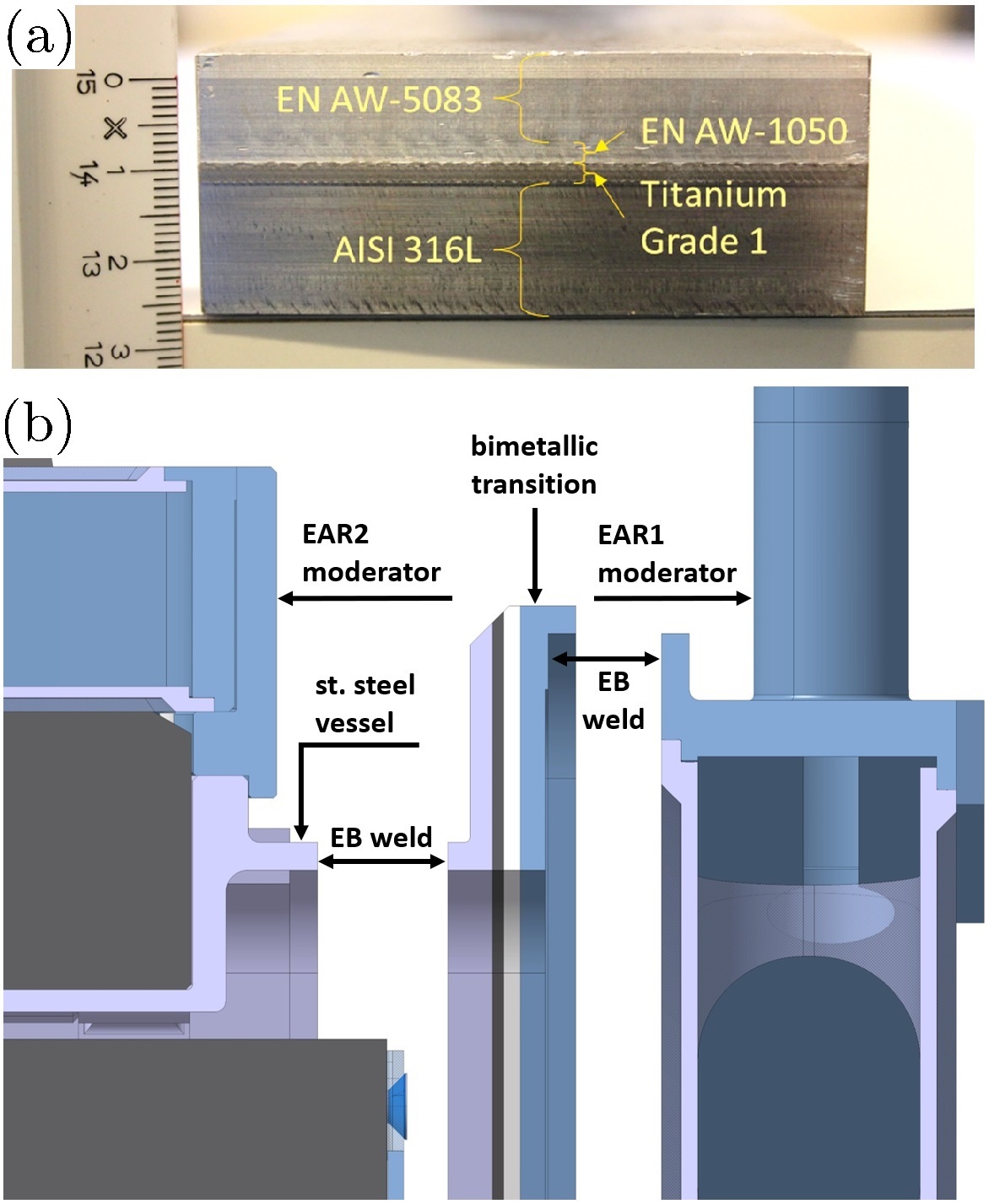}
    \caption{(a) Cross-cut of the explosive-bonded bimetallic transition between the stainless steel target vessel and the aluminum EN AW-5083 H112 EAR1 moderator. (b) Schematic view of the explosive-bonded vessel-moderator connection.}
    \label{3-3_bimetallic_transition}
\end{figure}

The EAR1 moderator, the larger of the two, was manufactured by deep-milling technology, which reduced the number of welds needed but required a rigorous production-process qualification to ensure the necessary dimensional tolerances. This moderator is an integral part of the vessel. The transition between the aluminum alloy of the moderator and the stainless steel vessel is a rectangular-shaped strip cut out from an explosive-bonded plate. The challenge of such a solution lies in the qualification of the bimetallic plates available from industry and on the design of the interfaces between the different parts to ensure the desired gas tightness for the entire lifetime of the target. The transition from aluminum alloy (EN AW-5083 H112) to stainless steel (AISI 316L) includes two additional ductile layers of pure aluminum (EN AW-1050A H14) and pure titanium (Grade~1) to ensure the good quality of the explosive bonding process (Fig.~\ref{3-3_bimetallic_transition}). To validate the quality of the bond in terms of impurities, voids, and mechanical resistance, CERN established strict specifications and quality controls: the bond was subjected to helium leak tests (max. leak rate accepted: 10$^{-7}$~mbar\,L/s), and a dye penetrant inspection was performed in accordance with the EN 10228\nobreakdash-2 standard on 100\% of the machined surfaces. Ultrasonic tests were carried out for each explosive-bonded interface according to EN~10288\nobreakdash-4, along with metallographic inspections.

\begin{figure}[h]
    \centering
    \includegraphics[width=\columnwidth]{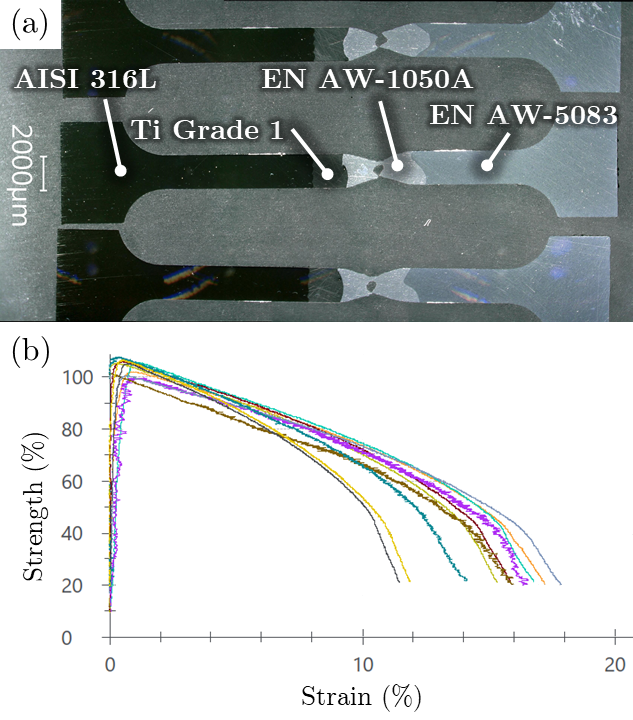}
    \caption{(a) Bimetallic-transition specimens tested in tension: for all of them, the interfaces between the different materials of the transition (stainless steel AISI 316L, titanium Grade~1, aluminum EN AW-1050A H14, aluminum EN AW-5083 H112) were stronger than the weakest base material (pure aluminum EN AW-1050A H14). (b) Stress--strain curves resulting from the tensile tests: the ultimate bond strength is 105~MPa.}
    \label{3-4_bimetallic_transition_tests}
\end{figure}

In addition to the nondestructive examinations, tensile tests and post-weld metallographic investigations were carried out on representative samples extracted from the original explosive-bonded plate. The tensile tests confirmed a bond strength of 105~MPa, higher than minimum acceptable value of 80~MPa and limited by the strength of the weakest base material, aluminum EN AW-1050A H14 (Fig.~\ref{3-4_bimetallic_transition_tests}). The post-weld metallographic investigations confirmed that, after electron-beam welding the explosive-bonded joint to the moderator, the Heat-Affected-Zone (HAZ) is confined far from the interfaces of the bimetallic transition within a safe distance of 3~mm, and thus it does not affect them.

A first prototype of the explosive-bonded joint, 3\nobreakdash-mm thick, failed the leak validation. A second prototype was built to find a safe thickness value for the joint: the prototype was designed to easily perform leak tests and reduce thickness by milling. After each leak test, the thickness of the prototype was reduced in steps of 5~mm, starting from 28~mm and down to 8~mm: no leak was detected at any iteration. It was concluded that, rather than its thickness, the main factor determining the soundness of a leak-tight joint is the choice of the original region of the plate from where the final joint is extracted. Regions that appear sound after preliminary dye-penetrant and ultrasonic inspections are the best candidates. The final product was selected on this basis, with a final joint thickness of 25~mm, the maximum permitted by assembly and space constraints.

\begin{figure}[h]
    \centering
    \includegraphics[width=\columnwidth]{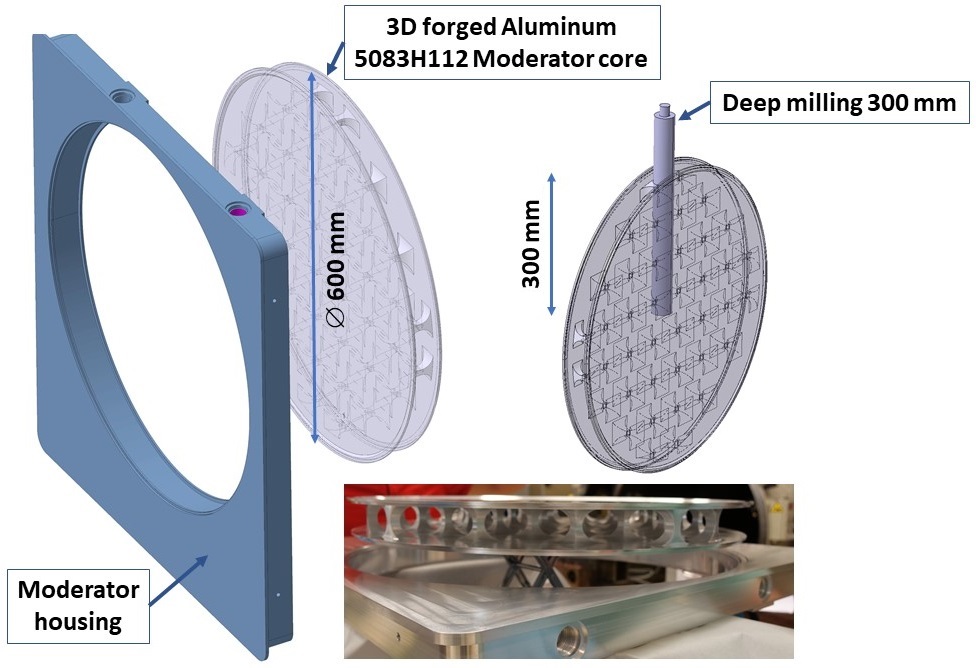}
    \caption{Schematic view of the deep milling procedure for the EAR1 moderator manufacturing. The central part is entirely machined from a single solid aluminum block (EN AW-5083 H112). It is then bonded to the housing via electron-beam welding.}
    \label{3-5_moderator_EAR1}
\end{figure}

An innovative manufacturing process was developed to accommodate both mechanical and physics requirements of the EAR1 moderator. Shaped like a short cylindrical vessel, the two flat end surfaces, each 3~mm thick, would exceed the maximum acceptable deformations and stresses induced by the 2.5~bar pressure without additional stiffening elements. The use of a precise deep-milling manufacturing technique offered the possibility to add such stiffeners while preserving the monolithic nature of the part and avoiding welds, which would increase failure risks. Up to 80\% of the material was removed from a solid aluminum disk, with the remaining 20\% shaped to obtain deformations and stresses compliant with the specification. The resulting monolithic center part of the moderator was then welded to the outer housing by electron-beam welding. The outer housing includes the interface with the explosive-bonded joint (Fig.~\ref{3-5_moderator_EAR1}).

Special effort was put into the selection process of the bulk material for the disk in terms of homogeneity and grain size, low number of defects, and impurity. The choice fell on 3D forged aluminum EN AW-5083 H112 blocks for its good compliance with the specifications and stability during machining. Compliance with the specifications was checked before machining by dye-penetrant and ultrasonic inspections. In addition, the qualification of the electron-beam welding process was carried out on representative samples.
To complete the qualification process of the EAR1 moderator, vacuum and pressure tests were successfully carried out to ensure the compliance with pressure vessel safety standards.


\section{\label{sec:CFD}Nitrogen cooling system: Computational Fluid Dynamics}

Computational Fluid Dynamics (CFD) studies were carried out to support and optimize the design of the target. The optimization consisted of maximizing the cooling efficiency (minimizing the maximum steady-state temperature) while minimizing the pressure drop and the locations of high flow velocity inside the target. The CFD calculations also provided input for the thermo-mechanical analyses (see Sec.~\ref{sec:FEM}) and for the design of the cooling station, (see Sec.~\ref{sec:nitrogen}).

The commercial CFD software \textsc{Ansys CFX}~\cite{ansys} has been used throughout the studies to model simultaneously the solid (lead slices) and the fluid bodies (nitrogen), and perform conjugate heat transfer calculations. The map of average beam power deposition, calculated with \textsc{FLUKA} Monte Carlo simulations, was imported as a power-generation boundary condition. The Shear Stress Transport (SST) turbulence model~\cite{Menter12} was employed to accurately compute the momentum and the thermal boundary layers.

\subsection{\label{sec:cfd:design}Design and performance optimization}

The CFD design optimization encompassed three main components: inlet and outlet ducts of the target vessel, cradle, and anti-creep plates.

The optimization was initially focused on tuning the number of inlets and outlets, their position, and dimensions. The final configuration presents the best compromise between the target pressure drop and a balanced distribution of nitrogen, while complying with the space constraints of the assembly.

\begin{figure}[h]
    \centering
    \includegraphics[width=\columnwidth]{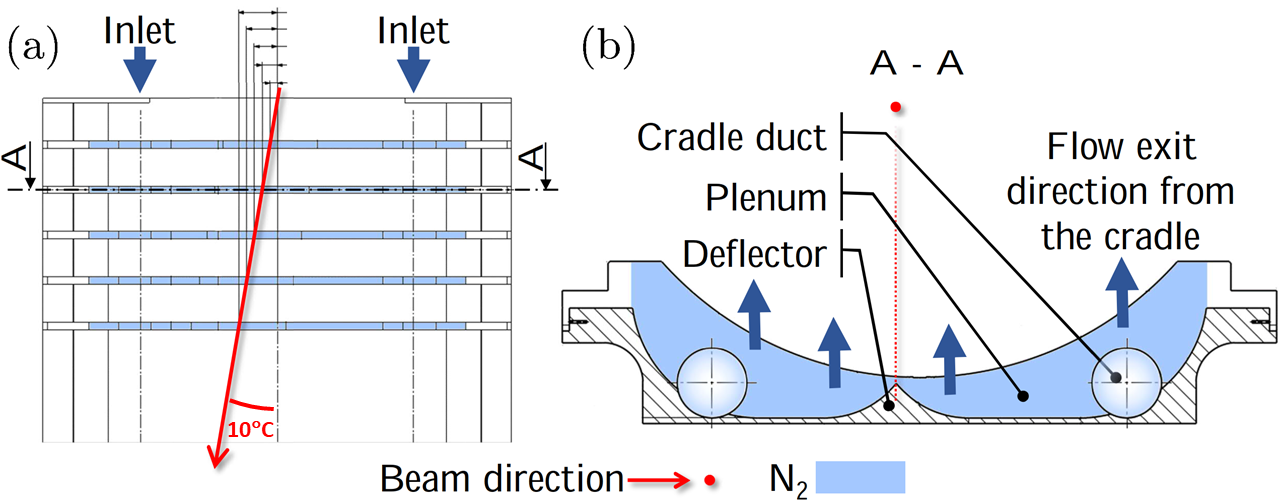}
    \caption{(a) Top view of the main core assembly (see also Fig.~\ref{3-1_target_assembly}). Nitrogen flows through the channels machined in the anti-creep plates, perpendicular to the view and toward the observer. The beam does not impact the lead surface perpendicularly, but at a 10$^\circ$ angle. (b) Cross-cut of the aluminum cradle through the second row of channels, showing the plenum volume and the geometry used to guide the flow into the channels.}
    \label{4-1_flow_in_cradle}
\end{figure}

The cradle was designed to evenly distribute the nitrogen flow from the inlets to the channels between the lead slices. Large plenum volumes, as shown in Fig.~\ref{4-1_flow_in_cradle}, induce a deceleration of the fast flowing nitrogen, while their curved geometry inhibits flow separation at the walls due to the adverse pressure gradient, typical of a sudden expansion. As shown in Fig.~\ref{4-1_flow_in_cradle}, the incident beam axis forms a 10$^\circ$ angle with the target to reduce the EAR1 background caused by $\gamma$-rays and high-energy charged particles \cite{Esposito20}. The beam is sufficiently diverging so that a dedicated dump is not required downstream the target in addition to the iron and marble already present around the target area (see Sec.~\ref{sec:RP}). The cradle design is complemented with flow deflectors aligned with the beam axis to focus the cooling flow on the hottest regions of the lead surface.

\begin{figure}[h]
    \centering
    \includegraphics[width=\columnwidth]{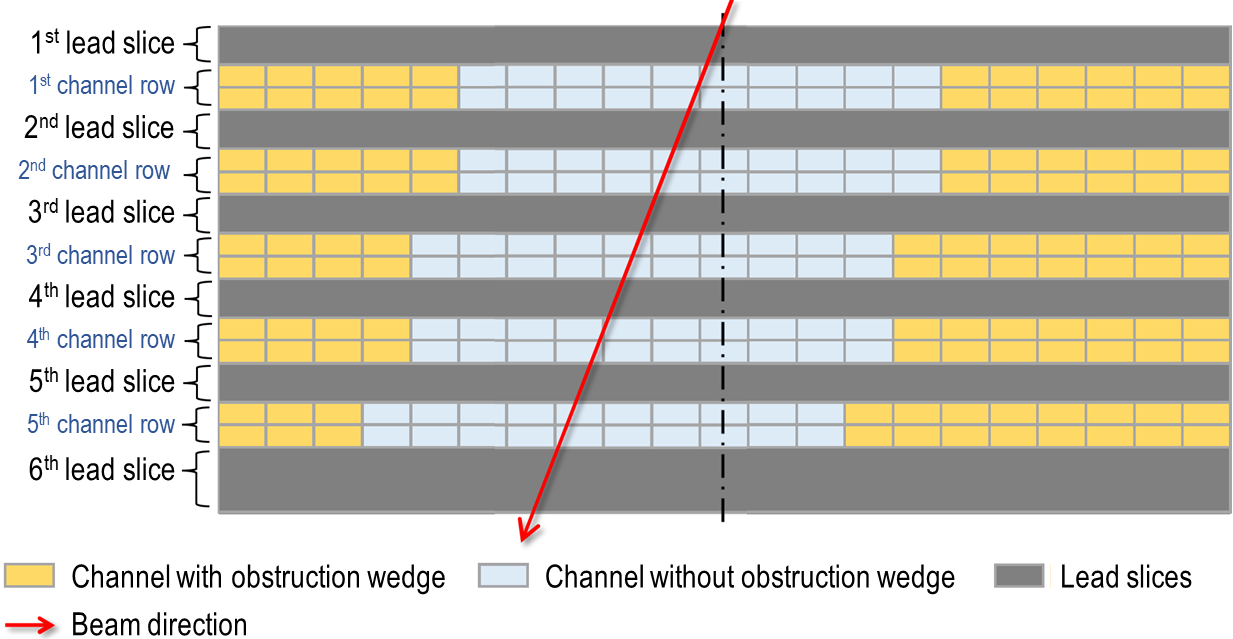}
    \caption{Schematic view of the main core assembly (see also Fig.~\ref{3-1_target_assembly}) showing, highlighted in yellow, the channels with wedge-shaped obstructions. The obstructions partially restrict the flow in the channels located farther from the beam axis, increasing the flow rate on the regions of lead surface directly impacted by the beam.}
    \label{4-2_obstructed_channels}
\end{figure}

\begin{figure}[h]
    \centering
    \includegraphics[width=\columnwidth]{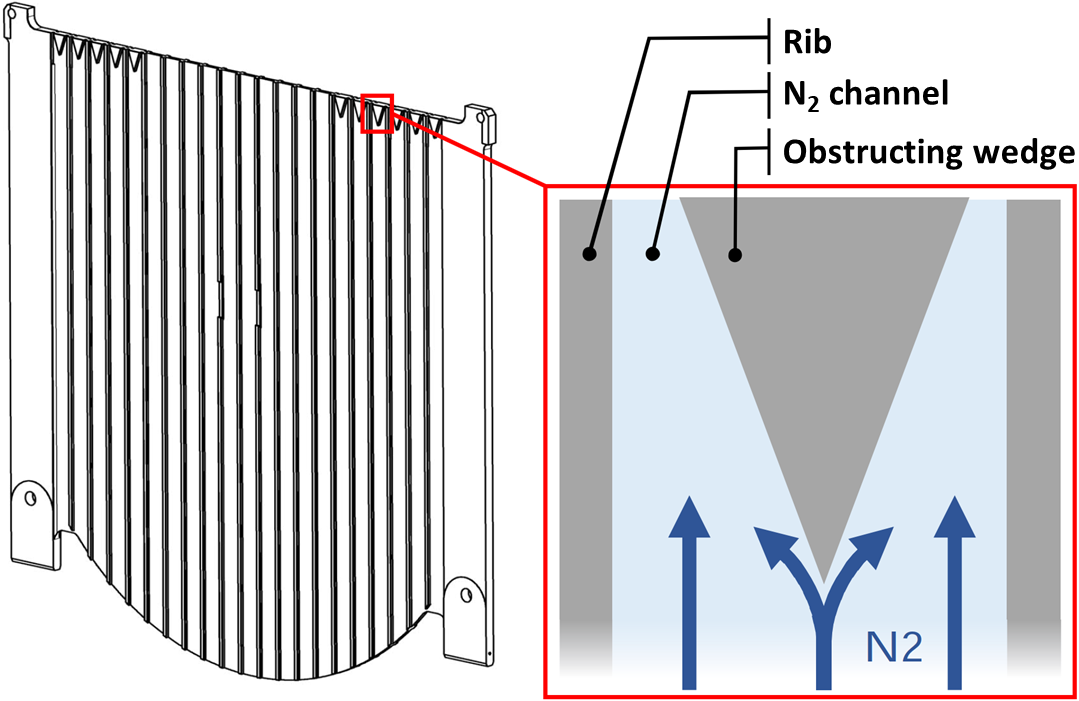}
     \caption{Example of an anti-creep plate and detail of the wedge-shaped obstructions. The obstructing wedges increase the pressure drop but help deliver higher flow rate on the regions of lead surface directly impacted by the beam.}
    \label{4-3_obstructing_wedge}
\end{figure}

The anti-creep plates drive the nitrogen over the lead surface while containing the creep deformation of the material. The distribution of flow rate among the channels is optimized by adding wedge-shaped obstructions in the channels that are located farther from the beam, as schematically shown in Fig.~\ref{4-2_obstructed_channels} and Fig.~\ref{4-3_obstructing_wedge}. The geometry of these obstructing wedges was carefully designed to minimize pressure losses and flow disturbances.

\subsection{\label{sec:cfd:performance}Cooling performance}

The cooling performance is presented for four operational cases, summarized in Table~\ref{T4-1_cfd_cases}. Cases A and B reflect the two normal operational limits of the cooling station, while cases C and D assess the cooling performance when some of the cooling channels are obstructed by lead due to creep deformation. The cases C and D are defined by considering no flow nor heat transfer in the channels that are adjacent to the lead surfaces where the highest temperature values are expected. The channels simulated as obstructed are the ones closest to the beamline, on both faces of the 2nd slice and on the downstream face of the 6th slice (according to the nomenclature in Fig.~\ref{4-2_obstructed_channels}). For each of these faces, three channels are considered obstructed in scenario C, while five channels are considered obstructed in scenario D. Given the expected creep deformations discussed in Sec.~\ref{sec:FEM}, these two cases represent rather conservative degraded scenarios.

\begin{table}[h]
    \caption{\label{T4-1_cfd_cases}Normal and degraded scenarios simulated for cooling performance assessment.}
    \begin{tabular}{ r c d }
    \hline\hline
    \thead{Case} & \thead{n$^\circ$ clogged\\channels} & \multicolumn{1}{c}{\thead[l]{Volume flow\\(Nm$^3$/h)}} \\
    \colrule
               \rule{0pt}{3mm}A &                                   0 &  800                                                   \\
               B &                                   0 & 1000                                                   \\
               C &                                   3 &  800                                                   \\
               D &                                   5 &  800                                                   \\
    \hline\hline
    \end{tabular}
\end{table}

\begin{figure}[h]
\centering
\includegraphics[width=\columnwidth]{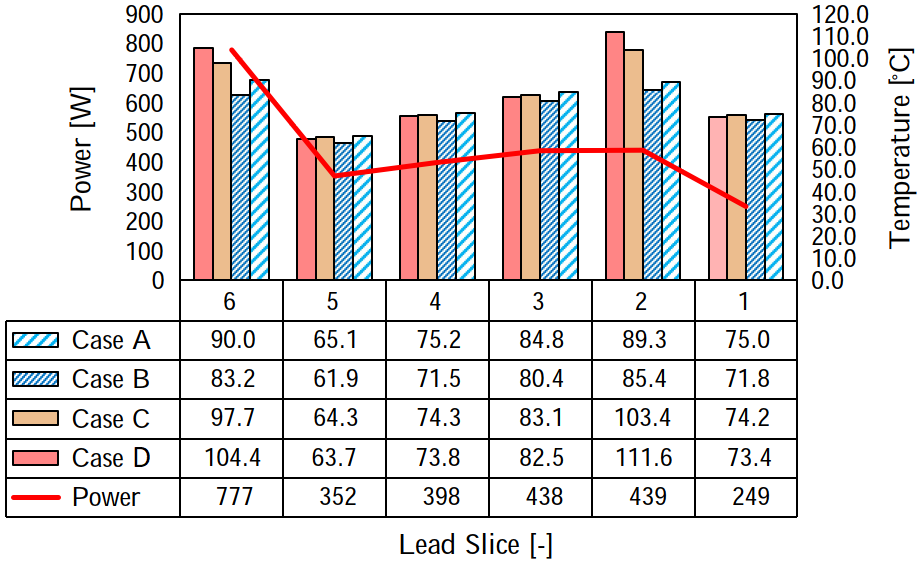}
\caption{Maximum steady-state temperature in each lead slice and scenario described in Table~\ref{T4-1_cfd_cases}, considering an average beam power of 5.4~kW and a beam size of 15~mm (RMS). The highest temperature is reached in the 2nd slice, followed by the 6th slice.}
\label{4-4_ss_max_temp}
\end{figure}

Considering a beam average power of 5.4~kW (2.7~kW absorbed by the lead slices) and a beam size of 15~mm (RMS), the maximum steady-state temperature in each slice is reported in Fig.~\ref{4-4_ss_max_temp}, and it appears correlated to the heat absorbed by each slice.
The 1st and 6th slices are an exception: the 1st slice absorbs the lowest amount of heat without being the coldest one, while the 6th slice absorbs the highest amount of heat without being the hottest one. This is due to the fact that they are cooled only on one of the two faces.

The maximum steady-state temperature values occur in the 2nd and 6th slices, ranging between 85--89$^\circ$C and 83--90$^\circ$C, respectively, and depending on the nitrogen flow. These values are likely to increase if substantial creep occurs where lead is subjected to higher temperature and stress values, with the 2nd slice experiencing the highest increase (25\% for case D).

\begin{figure}[b]
\centering
\includegraphics[width=\columnwidth]{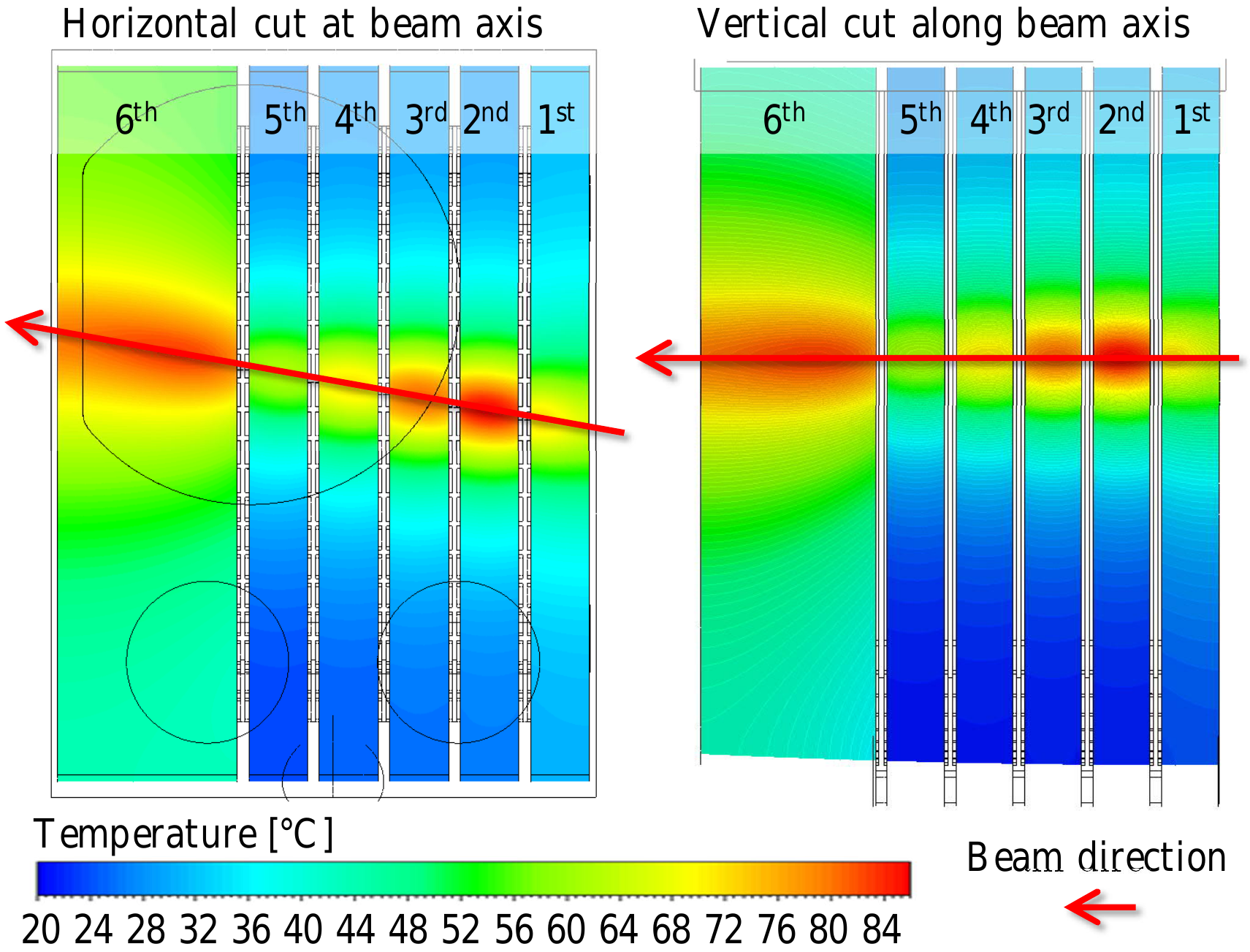}
\caption{Steady-state temperature distribution in the lead slices, cross-sections intersecting the beam axis. The beam axis is not orthogonal to the lead surface, but tilted by 10$^\circ$ in the horizontal plane to reduce the EAR1 background caused by $\gamma$-rays and high-energy charged particles.}
\label{4-5_ss_temp_contour}
\end{figure}

\begin{figure}[h]
\centering
\includegraphics[width=\columnwidth]{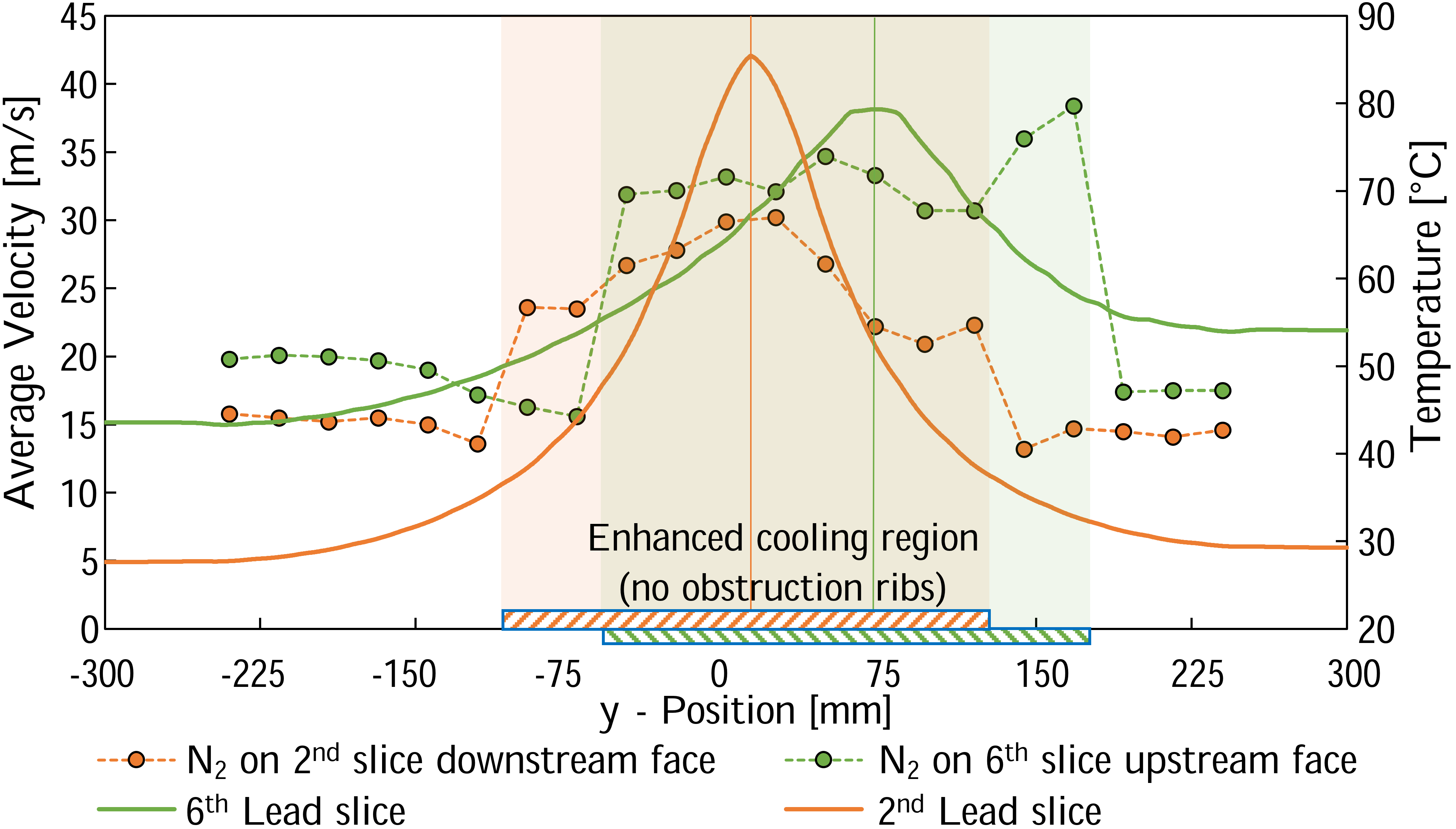}
\caption{Steady-state temperature profile along the horizontal at the height of the beam axis, for the downstream and upstream face of the 2nd and 6th slices, and average nitrogen velocity in the corresponding cooling channels.}
\label{4-6_ss_temp_flow_channels}
\end{figure}

Within each slice, the temperature is highest close to the beam axis, as shown in Fig.~\ref{4-5_ss_temp_contour}.
For the downstream and upstream face of the 2nd and 6th slices, respectively, the temperature profile along the horizontal at the height of the beam axis is shown in Fig.~\ref{4-6_ss_temp_flow_channels}. The average velocity of the nitrogen flow in the corresponding channels is also represented: the speed is higher in the channels without obstructing wedges, which translates into a higher heat transfer coefficient (HTC) where most of the beam power is absorbed.

The HTC values were estimated using the Gnielinski formulation with entrance length correction~\cite{Gnielinski90}, using the velocity and surface temperature values obtained from the CFD simulations, and the fluid properties at the bulk temperature of nitrogen. The HTC values computed analytically at beam height exceed the ones from the CFD simulations by 10--30\%, which is a reasonable match given the complexity of the geometry.

The pressure drop estimated by simulations and the peak value of velocity in the target for the two operational cases are summarized in Table~\ref{tab:cfd:press}.

\begin{table}[h]
    \caption{\label{tab:cfd:press}Nitrogen pressure drop in the target and peak velocity values for the four scenarios described in Table~\ref{T4-1_cfd_cases}.}
    \begin{tabular}{ r d d d d }
    \hline\hline
    & \multicolumn{1}{c}{\thead{A}} & \multicolumn{1}{c}{\thead{B}} & \multicolumn{1}{c}{\thead{C}} & \multicolumn{1}{c}{\thead{D}} \\
    \colrule
       \rule{0pt}{3mm}Pressure drop (kPa) &  4.7 &  7.0 & 4.8 & 5.0 \\
    Maximum velocity (m/s) & 77.2 & 89.5 &     &     \\
    \hline\hline
    \end{tabular}
\end{table}

The pressure drop in the target is expected to be between 4 and 7~kPa depending on the flow. A detailed analytical estimation of the pressure drop in the target vessel was calculated via the loss coefficient method using coefficients from~\cite{Rohsenow98,Idelchik86,ASHRAE09}, and a good agreement (difference lower than 10\%) was found when compared to simulations.
The blockage of cooling channels due to creep, as in case D, results in an increase of 6\% in pressure drop, which is well within the range in which the cooling station can operate.

The analysis also shows that the nitrogen velocities in the target are below Mach~0.3, thus the flow can be considered incompressible everywhere, with the highest velocities occurring at the bend of the vessel inlets. Inside the target vessel, specifically in the cradle ducts, the nitrogen flows relatively fast, decelerating as it flows into the plenums along the ducts. The analysis results reveal that velocities up to 65~m/s are expected to occur at the entrance of a few channels due to the apparent cross section reduction and flow separation at the entrance of the channels. The nitrogen flowing through the narrow sections beside the obstruction wedges can reach velocities between 70~m/s and 87~m/s for cases A and B, respectively; however, the velocities at the beam impinging spots are below 40~m/s. The highest velocity values, besides taking place in limited and localized areas over colder lead surfaces, are within the acceptable threshold value to avoid erosion of 110~m/s (according to~\cite{API13}).


\section{\label{sec:FEM}Thermo-mechanical assessment of the lead core} 

The time evolution of temperature, stress, and strain in the target were estimated by means of CFD and FEM analyses with \textsc{Ansys} and \textsc{LS-DYNA}~\cite{ansys}.

\subsection{\label{sec:FEM:thermal}Thermal analyses}

The temperature dependence of every material property has been included in the analyses. The reference sources for the thermal properties are~\cite{Feder58,Ho72,Bonnie93}.

The most conservative assumption that can be made regarding the beam load is that all of the six pulses hit the target consecutively with the minimum period of 1.2~s. The six pulses are then followed by a cool-down of 30~s, totaling a cycle of 36~s (hereinafter referred to as \textit{supercycle}). The load boundary condition for the thermal analyses is a power generation field in the target, resulting from the interaction between beam and target and obtained by means of \textsc{FLUKA} Monte Carlo simulations.

The cooling effect of nitrogen is modeled by a convective boundary condition, where the heat transfer coefficient (HTC) field was imported from the CFD simulations described in Sec.~\ref{sec:CFD}. The average HTC is 63.8~W\,m$^{-2}$\,K$^{-1}$, but the flow is optimized to be faster at the beam impinging spot, where the HTC reaches a value of 130~W\,m$^{-2}$\,K$^{-1}$.

\begin{figure}[h]
    \centering
    \includegraphics[width=\columnwidth]{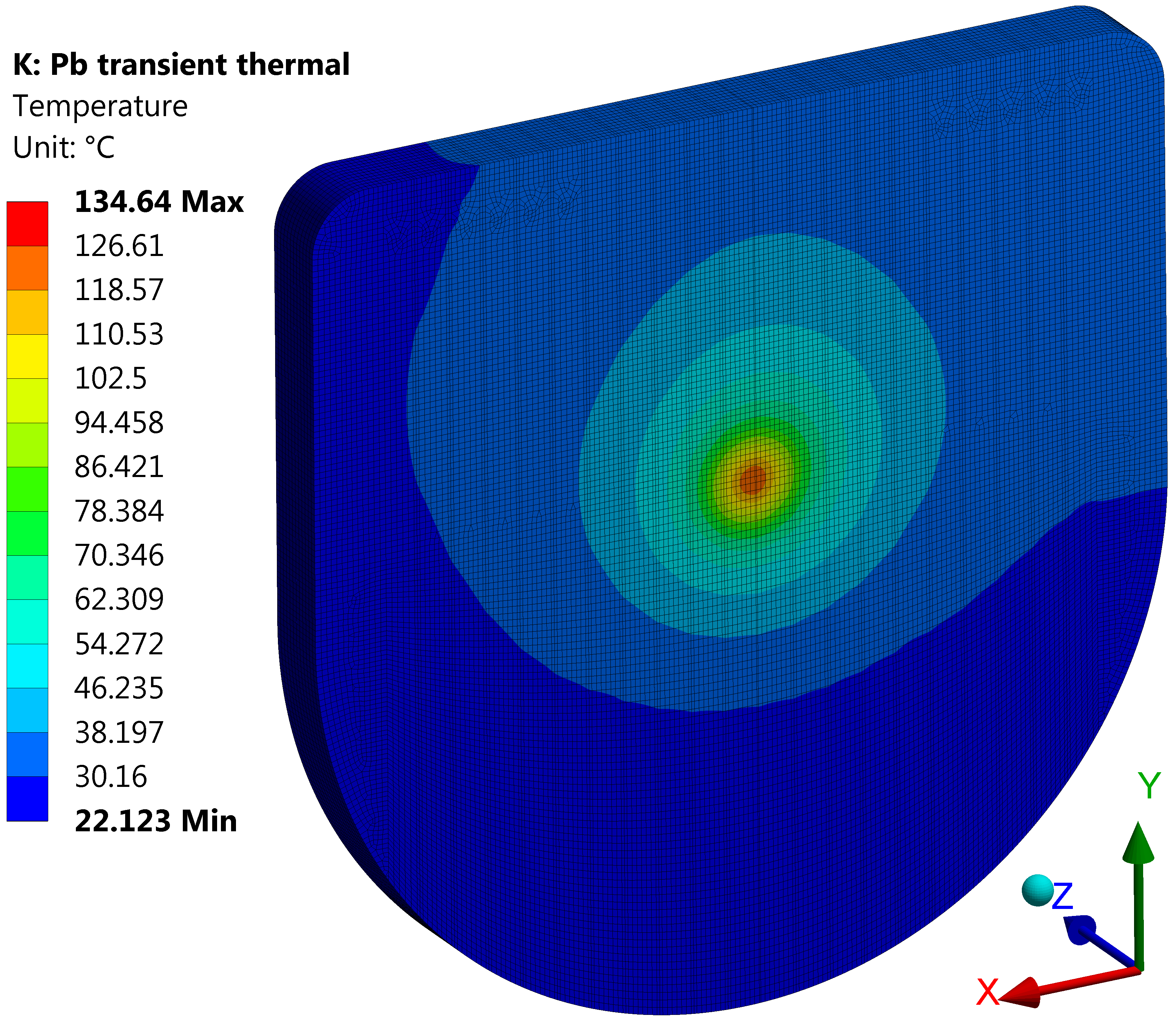}
    \caption{Temperature map, in the 2nd lead slice crossed by the beam, at the instant of peak temperature. The peak temperature is equal to 135$^\circ$C.}
    \label{5-1_Pb_temp_contour}
\end{figure}

\begin{figure}[h]
    \centering
    \includegraphics[width=\columnwidth]{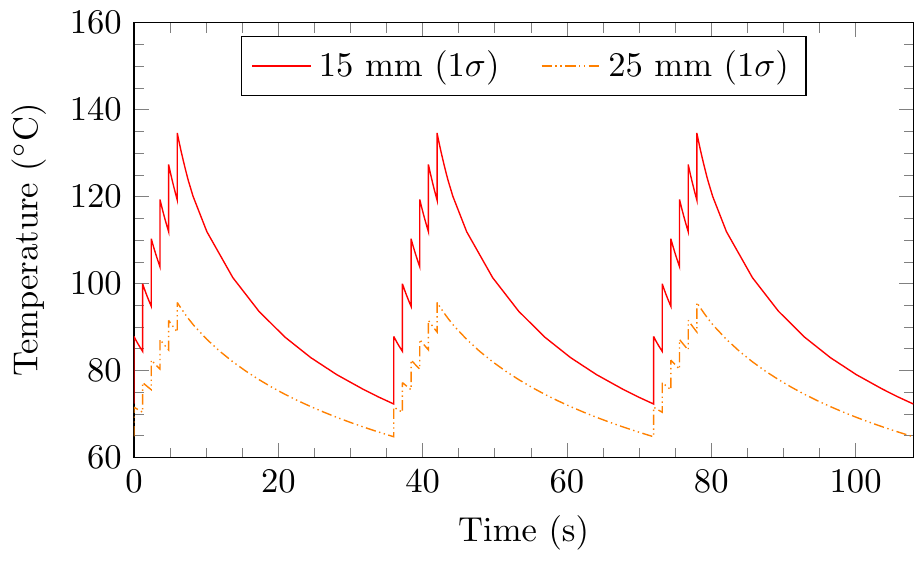}
    \caption{Maximum temperature in the target Pb core during three 36-s supercycles once periodic steady state is reached, with beam size of 15~mm and 25~mm (RMS).}
    \label{5-2_plot_Pb_temp}
\end{figure}

The 2nd lead slice crossed by the beam is subject to the highest energy deposition and temperature. The results of thermal simulations for this slice are shown in Fig.~\ref{5-1_Pb_temp_contour} and \ref{5-2_plot_Pb_temp}: Fig.~\ref{5-1_Pb_temp_contour} shows the temperature field at the instant of peak temperature, while Fig.~\ref{5-2_plot_Pb_temp} shows the plot of maximum temperature vs. time during three 36\nobreakdash-s supercycles, once periodic steady state is reached. The peak temperature is 135$^\circ$C, safely lower than the melting point of lead (327$^\circ$C) but high enough to significantly decrease strength and promote creep. Figure~\ref{5-2_plot_Pb_temp} also shows the benefits of a change of optics, if feasible: a larger beam size (25~mm RMS) lowers the peak temperature to 96$^\circ$C.

A series of six radiation-hard thermocouples have been placed inside the target, directly in contact with the lead slices. Although they can provide benchmarks for the thermal simulations and feedback on the status of the target core, they are not essential for the target operation in case they fail prematurely.

\subsection{\label{sec:FEM:stress}Stress analyses}

The target is subject to mechanical stress due to two effects, both included in the analyses: quasi-static stresses and dynamic stresses. Quasi-static stresses are induced by heat from the absorbed beam power in the target, which provokes temperature and thermal expansion gradients. Furthermore, each beam pulse is absorbed in 7~ns (RMS), provoking an impulsive temperature rise and consequent thermal expansion, which causes the propagation of stress waves and induces dynamic stresses.

The properties used for the structural analyses are taken from~\cite{Hofmann70,Rack78,Frost82}. Two constitutive models were used to model plasticity: pure isotropic hardening (without Bauschinger effect), and pure kinematic hardening (with Bauschinger effect)~\cite{Lemaitre90}. Material characterization activities were carried out to develop a nonlinear kinematic hardening model for plasticity of pure lead. The details of this research will be reported in a dedicated publication.

\begin{figure}[h]
    \centering
    \includegraphics[width=\columnwidth]{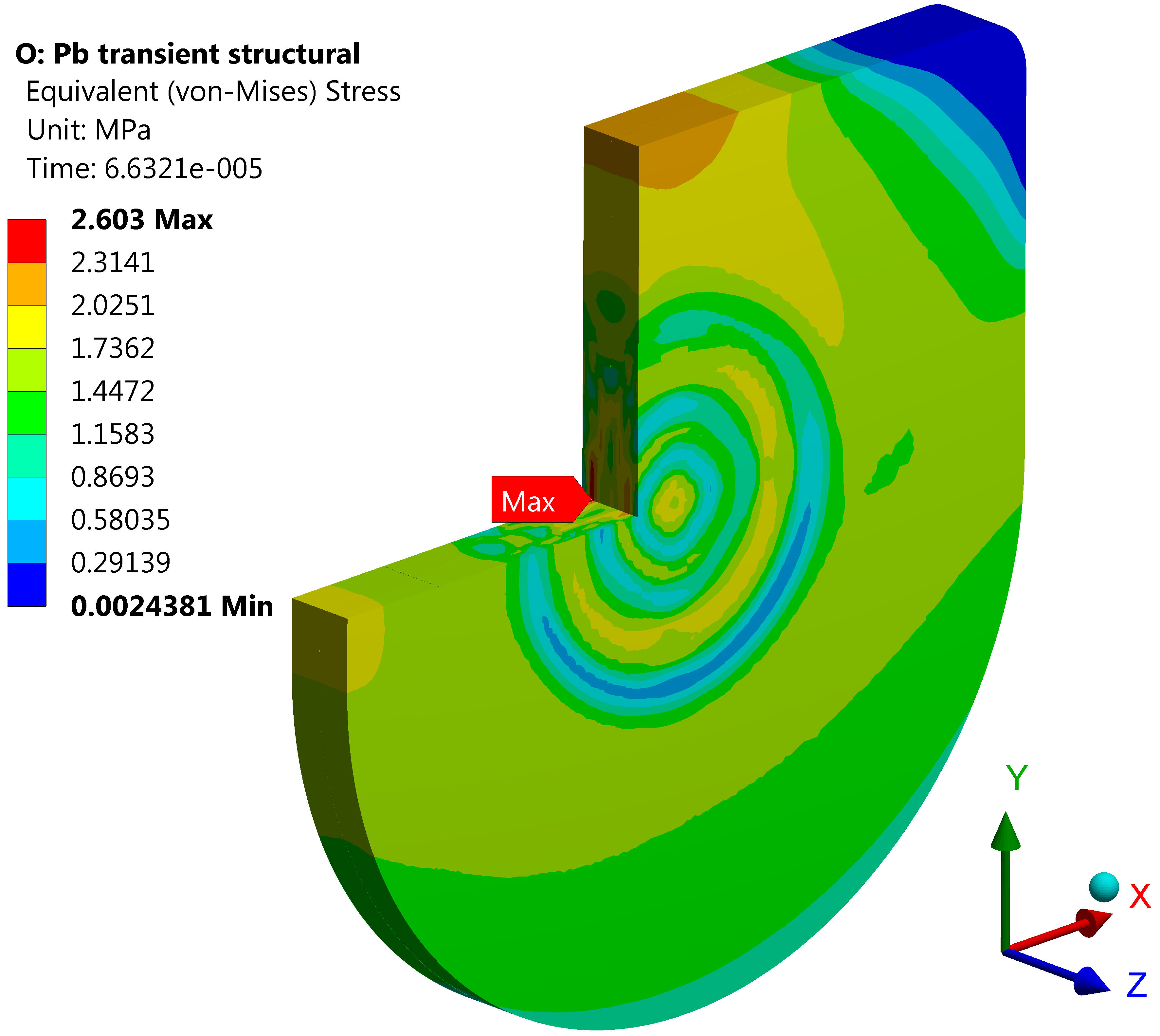}
    \caption{Peak von Mises equivalent stress in the target lead core during propagation of stress waves. The stress waves start propagating radially from the area of impact on the proton beam axis.}
    \label{5-3_Pb_stress}
\end{figure}

\begin{figure}[h]
    \centering
    \includegraphics[width=\columnwidth]{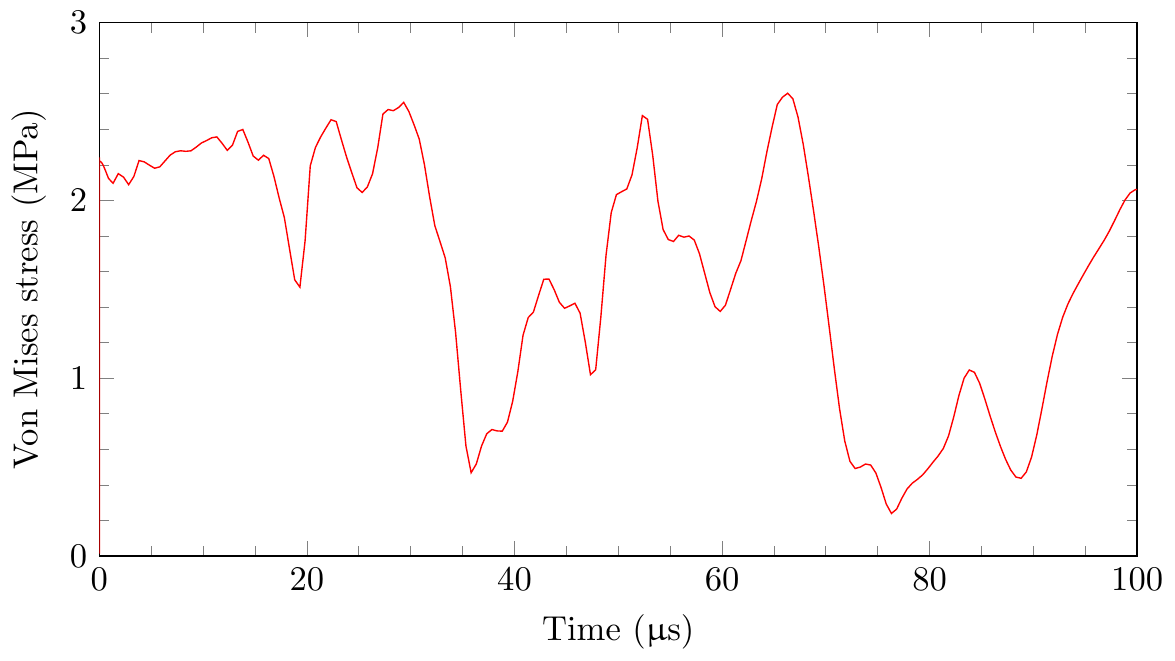}
    \caption{Von Mises stress vs. time in the target Pb core after the sixth (and last) pulse of a supercycle. The oscillations are provoked by the instantaneous temperature spike due to heat dissipation of the beam pulse.}
    \label{5-4_plot_Pb_stress}
\end{figure}

Figures~\ref{5-3_Pb_stress} and \ref{5-4_plot_Pb_stress} show some of the results of the structural simulations featuring the pure kinematic hardening model. Figure~\ref{5-3_Pb_stress} shows the von Mises equivalent stress map during the stress-wave propagation stage. Figure~\ref{5-4_plot_Pb_stress} shows the oscillations, due to dynamic effects, of the von Mises equivalent stress after the sixth (and last) pulse in a supercycle. The peak value of von Mises stress is 2.6~MPa, while pressure and maximum principal stress reach 80~MPa: the hydrostatic component of stress is dominant. The stress level is quite high if compared to the strength of pure lead (plastic flow begins below 1~MPa); thus, a target prototype was tested under beam irradiation in the HiRadMat facility at CERN~\cite{Efthymiopoulos11}. The outcome of the beam irradiation test was positive as no defects and internal voids were observed inside the material by neutron tomography inspections~\cite{Esposito20}. Further details of this beam irradiation test will be presented in a separate publication.

\subsection{\label{sec:FEM:creep}Creep analyses}

Pure lead is strongly affected by creep, i.e. plastic flow of material with time, especially at the temperature reached in the n\_TOF target (135$^\circ$C max.). Creep phenomena could induce lead to flow into some of the cooling channels, obstructing them and reducing the cooling efficiency. The creep strain rate for each level of temperature and stress is plotted in the \textit{deformation-mechanism maps}~\cite{Frost82}. A constitutive model able to reproduce lead creep over all the stress and temperature ranges of interest for the n\_TOF target is not readily available in \textsc{Ansys}; thus, a new model was coded as User Programmable Feature (UPF). The new model consists of a series of piecewise functions, each composed of multiple power laws (of the kind $\dot{\epsilon}=c\,\sigma^n$) with different coefficients: each piecewise function reproduces the deformation-mechanism maps for a single temperature; the creep strain rates for intermediate temperatures are obtained by logarithmic interpolation.

\begin{figure*}[t]
    \centering
    \includegraphics[width=1.6\columnwidth]{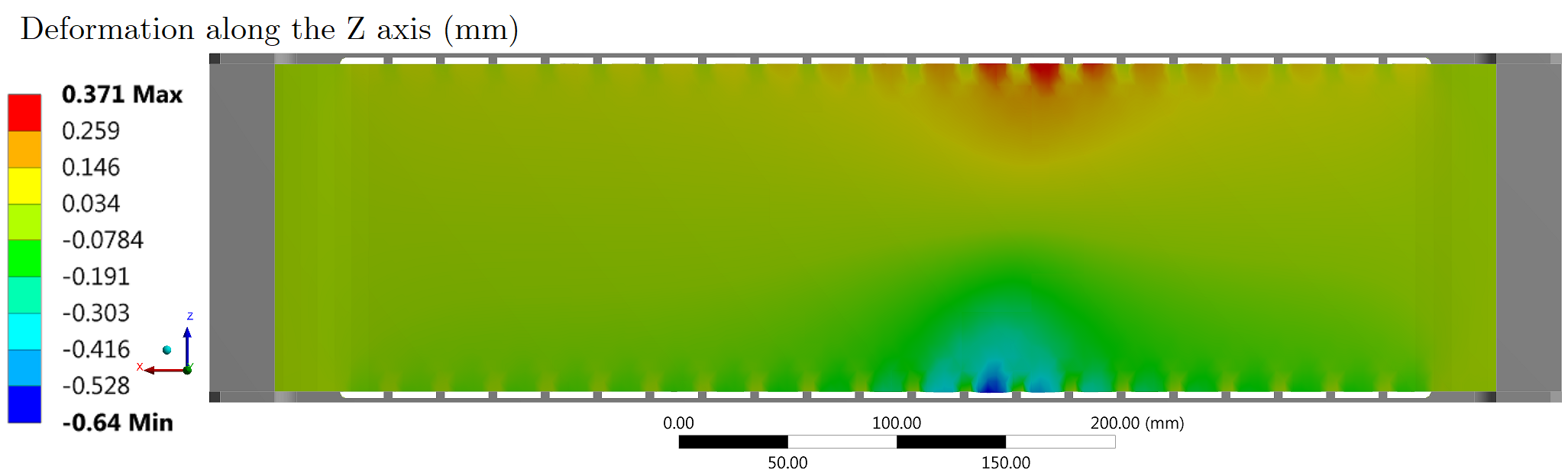}
    \caption{Penetration, due to creep, of the 15-cm thick lead slice into the cooling channels (view from the top, cross-section cut through the point of maximum penetration). The gray parts are the aluminum (EN AW-6082 T6) anti-creep plates with the cooling channels. The color map shows the maximum projected displacements after 2$\times$10$^8$~s (approximately 6 years and 4 months of continuous operation, twice the target lifetime).}
    \label{5-5_Pb_creep}
\end{figure*}

The sixth (and last) lead slice intersected by the beam is the thickest (15~cm) and is subjected to the largest creep deformations. A simulation, featuring the new creep constitutive model, was performed to estimate the maximum penetration, due to creep, of the lead slice into the cooling channels. The loads considered in the analysis are the weight of the slices and the temperature field at the instant of peak temperature. The temperature field is conservatively assumed to be constantly applied for 2$\times$10$^8$~s (approximately 6 years and 4 months of continuous operation, twice the target lifetime). The resulting displacements are shown in Fig.~\ref{5-5_Pb_creep}: the maximum penetration is 0.64~mm, whereas the channel depth is 3~mm.

As described in Sec.~\ref{sec:CFD}, degraded scenarios including the possibility of obstructed channels were considered in CFD simulations. The results of this analysis shows that the target operation would not be compromised: the lead slice would be subject to a maximum peak temperature of 160$^\circ$C, which is considered acceptable for the degraded scenario under examination.


\section{\label{sec:nitrogen}Design of the nitrogen cooling station}

The nitrogen gas for the target cooling system is produced by a nitrogen generator, recirculated in the cooling circuit, and cooled by chilled water. The design requirements of the nitrogen cooling system are summarized in Table~\ref{tab:cooling_requirements}. The nitrogen flow is filtered by High Efficiency Particulate Air (HEPA) and active carbon filters to capture particles and volatile compounds released in the nitrogen stream. The part of the cooling system between the target vessel and the filter casing containing the filters is enclosed in a sealed room under dynamic confinement: an extractor ensures that the pressure in the sealed room is lower than the surrounding environment. In case of a leak of activated nitrogen from the filter casing, the radioactive gas is therefore contained in the sealed room and extracted to a dedicated ventilation system in the target tunnel area. The maximum leak rate in the unconfined part of the circuit that has been considered acceptable after radiation protection studies (see Sec.~\ref{sec:RP}) is 5~L/h.

\begin{table}[h]
    \caption{\label{tab:cooling_requirements}Nitrogen cooling station: cooling requirements.}
    \begin{tabular}{ r r@{ }l }
    \hline\hline
             \thead[r]{Quantity} & \multicolumn{2}{c}{\thead{Value}} \\
    \colrule
    \rule{0pt}{3mm}Cooling power & 2.6 & kW                 \\
               Nominal flow rate & 780 & Nm$^3$/h           \\
                   Pressure drop & 30  & mbar               \\
          Gas supply temperature & 20\,$\pm$\,2 & $^\circ$C \\
                Maximum pressure & 500 & mbar               \\
                  \hline\hline
    \end{tabular}
\end{table}

\begin{figure*}[t]
    \centering
    \includegraphics[width=\textwidth]{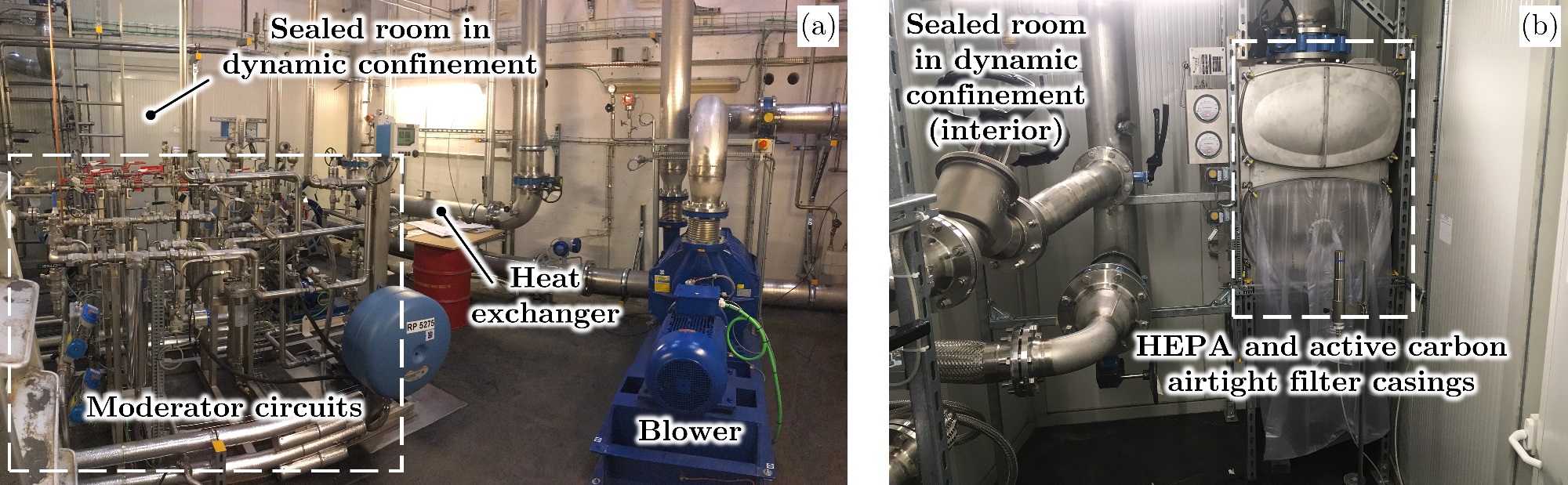}
    \caption{(a) Main components of the cooling and moderator station. The cooling nitrogen is compressed in the blower, cooled down in the heat exchanger, and filtered upstream and downstream the target. The station also includes the two seperate moderator circuits for EAR1 and EAR2, which can work independently with demineralized water or borated water. The filters are located inside an airtight room in dynamic confinement. (b) Photo inside the airtight room in dynamic confinement. The HEPA and active-carbon filter capture particles and volatile compounds in the nitrogen stream.}
    \label{6-1_cooling_station}
\end{figure*}

The cooling station supplies nitrogen gas to cool the target lead slices via AISI 304L stainless steel pipes. The main elements of the cooling station are shown in Fig.~\ref{6-1_cooling_station}.

The gas compressor (blower) is a multistage centrifugal compressor with a flow rate of 1000~Nm$^3$/h and a pressure rise of 285~mbar. It integrates features to minimize gas leaks such as magnetic coupling between compressor shaft and motor, silicone groove O-rings between casing discs, and EPDM seals at all bolted flanges. The 29\nobreakdash-kW heat exchanger is installed downstream of the blower to cool the hot nitrogen gas down to 20$^\circ$C via chilled water. It is of tubular type to minimize the pressure drop and has tubes arranged in horizontal configuration. The nitrogen stream is filtered by a HEPA filter upstream of the target and by a double-stage nuclear-grade HEPA plus active-carbon filter downstream of the target. Class 3 (ISO 10648\nobreakdash-2) gas-tight filter casings host all of the filters. The casing design allows the filter status to be monitored and the filters to be replaced without contact with nitrogen. The HEPA filters capture particles and aerosols with an efficiency greater than 99.98\% for a minimum particle size of 0.15~\textmu m. The active carbon filter captures any volatile compound released in the nitrogen gas stream. A gas-analysis system based on quadrupole mass spectrometry analyzes continuously the gas composition upstream and downstream of the double-stage filter casing with the HEPA and the active carbon filters. As previously mentioned, the filters are installed inside a sealed room kept in dynamic confinement via an extraction fan.


\section{\label{sec:RP}Radiation protection considerations}

The radiation protection (RP) implications of the target were thoroughly analyzed during its design phase; the outcomes of this analysis, which also took into account the return on experience from ten years of operation of Target~\#2, contributed to the optimization of the target design.
The main RP aspects addressed in the target design study are: stray radiation in accessible areas and compliance with CERN’s radiological area classification; critical aspects of the new cooling system; air activation in the target area and atmospheric releases; radioactive waste considerations in view of the target final disposal at the end of its lifetime.

\begin{figure}[h]
    \centering
    \includegraphics[width=\columnwidth]{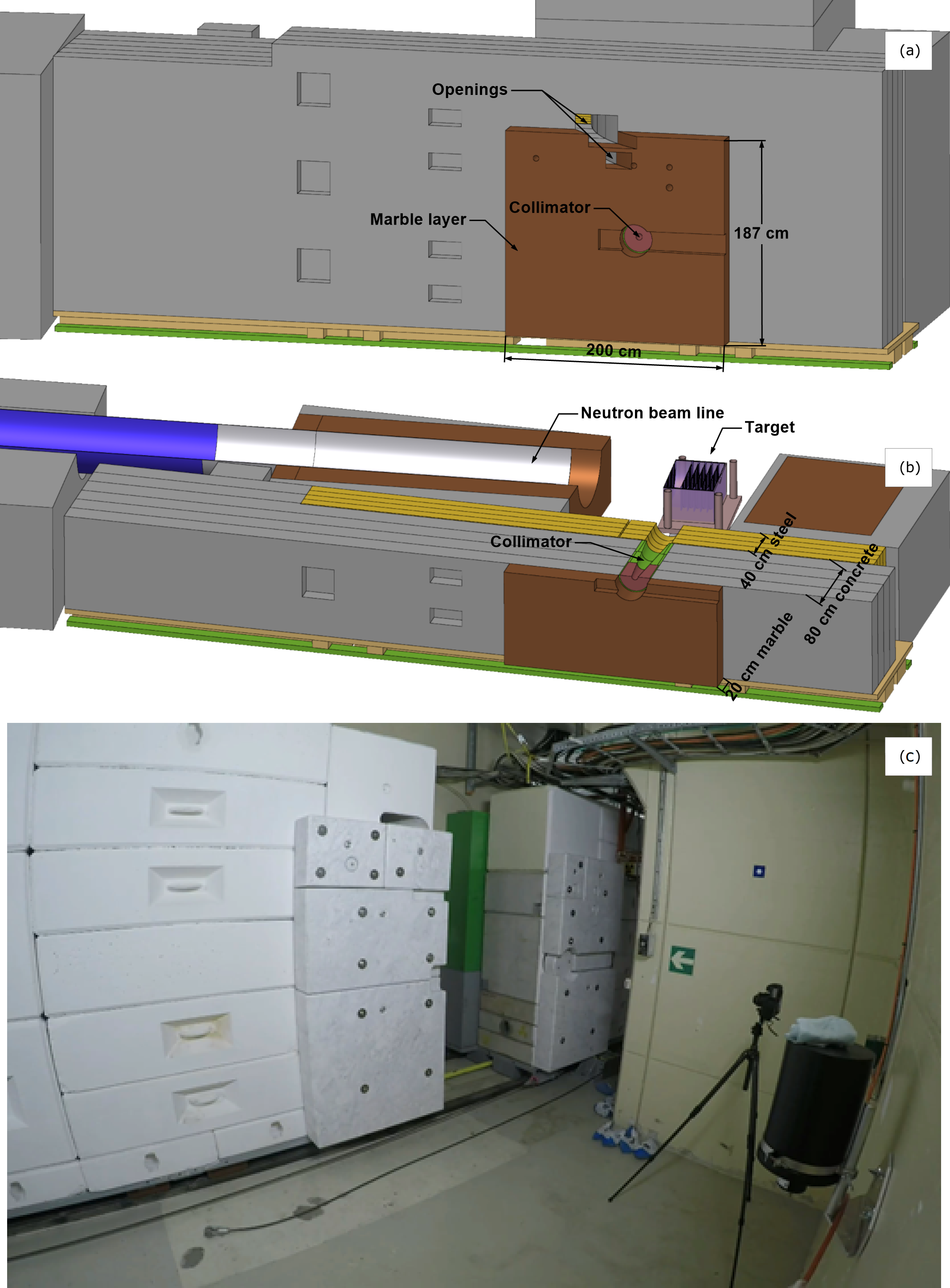}
    \caption{(a) 3D view of the \textsc{FLUKA} geometry of the new shielding. (b) 3D cross-sectional view inside the shielding wall. (c) Photo of the shielding in 'open' position: it can be moved on rails to access the target area.}
    \label{7-1_3D_mobile_shielding}
\end{figure}

Along with the target, the target shielding was also upgraded. The Target~\#2 shielding, composed of fixed concrete blocks with a maximum thickness of 200~cm, has been replaced with a new mobile version. The new shielding is composed of a first layer of 40\nobreakdash-cm thick iron followed by 80~cm of concrete and finally a 20\nobreakdash-cm layer of marble. The shielding wall has two main goals: to limit air activation in the target area during beam operation and subsequent releases into the environment (see Sec.~\ref{sec:RP:airact}) and to reduce the residual activation levels in the area during access. The new mobile shielding provides direct access to the target and enhances the potential applications of the facility (e.g., to perform irradiation of material samples in the proximity of the target~\cite{NEAR}). Several openings are present in the shielding to permit the future installation of a pneumatic system to transport radioactive samples and to access the target with endoscopes and radiation probes.
The shielding plays a major role in ensuring adequate radiation protection and, therefore, it was studied and optimized by means of \textsc{FLUKA} simulations (Fig.~\ref{7-1_3D_mobile_shielding}).

\subsection{\label{sec:RP:stray}Stray radiation}

\begin{figure}[h]
    \centering
    \includegraphics[width=\columnwidth]{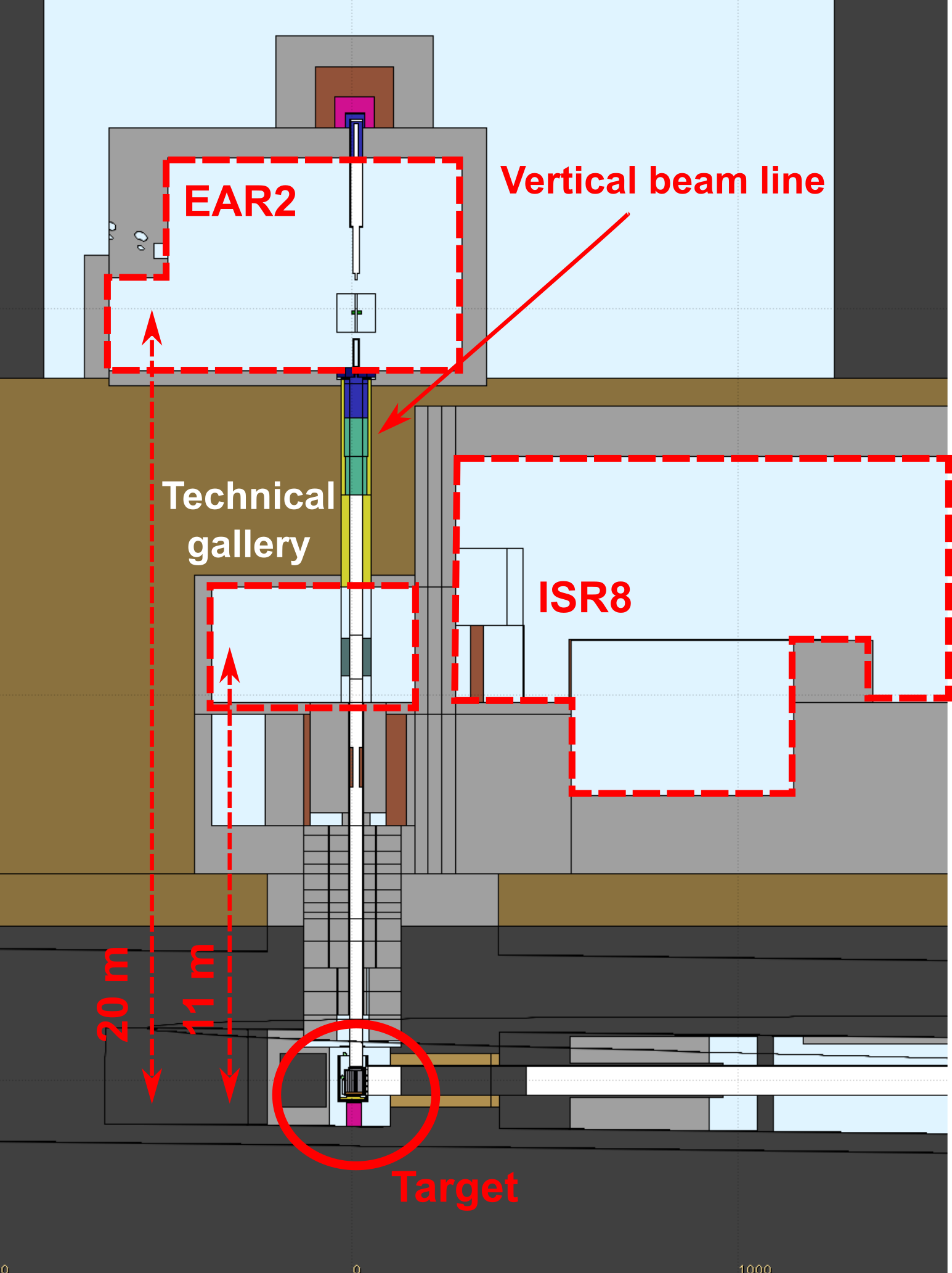}
    \caption{\textsc{FLUKA} geometry: side view of of the vertical beam line. The most important locations impacted by stray radiation are the experimental area EAR2 and the ISR8 area.}
    \label{7-2_2D_vertical_beam_line}
\end{figure}

As already mentioned in Sec.~\ref{sec:physics:neutron}, Target~\#3 provides, when compared to Target~\#2, a neutron fluence towards EAR2 twice as high in the energy range 100~keV--200~MeV; the locations of the n\_TOF facility that could be impacted by the resulting stray radiation are the ones surrounding the 20\nobreakdash-m vertical beam line, i.e. EAR2 and ISR8 (the eighth octant of the former CERN Intersecting Storage Ring, Fig.~\ref{7-2_2D_vertical_beam_line}).

The ISR8 area is shielded with concrete blocks and houses the cooling station and a space for storage.

\begin{figure}[h]
    \centering
    \includegraphics[width=\columnwidth]{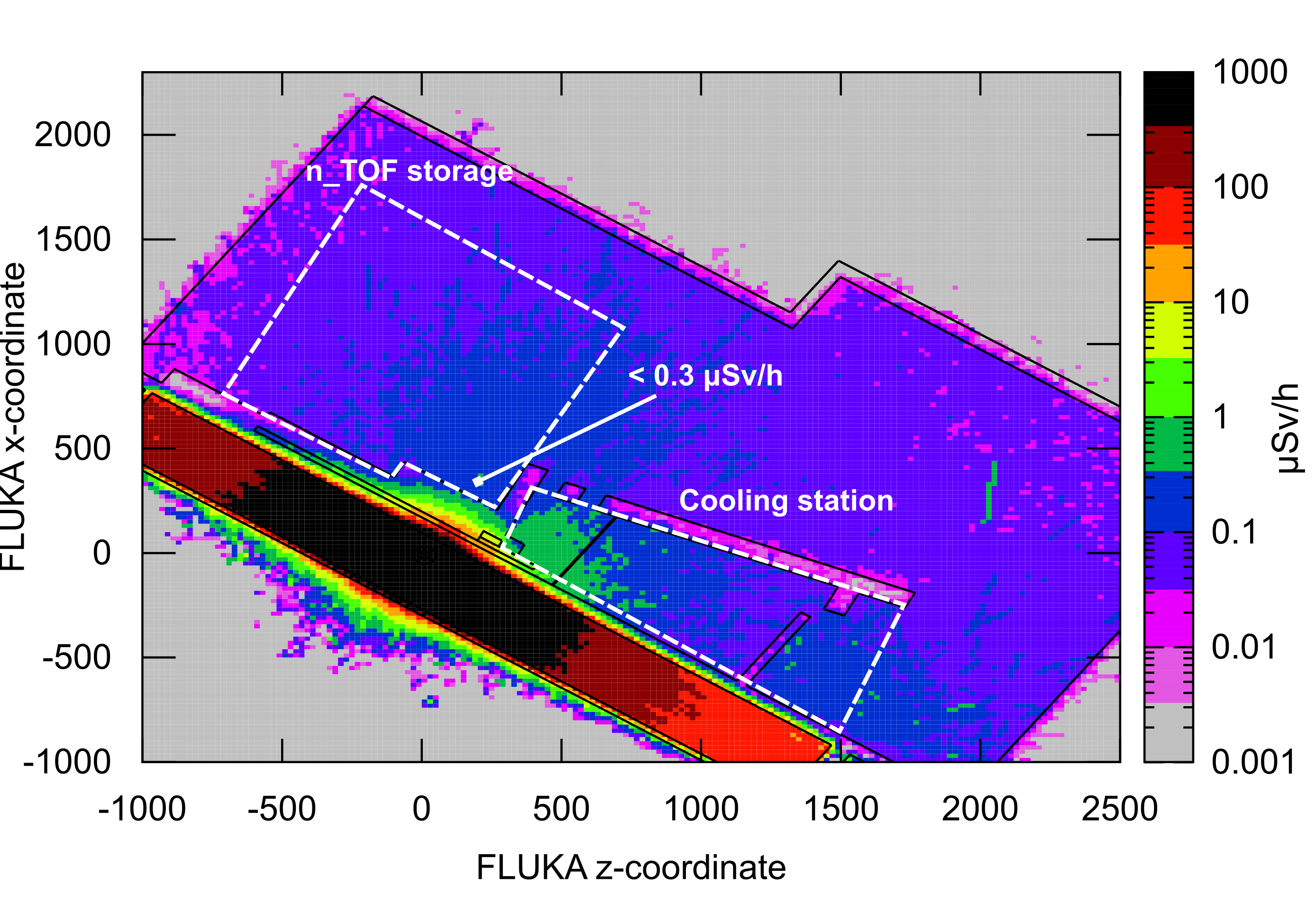}
    \caption{Prompt H*(10) rate in ISR8 for the maximum average beam intensity of 1.67$\times$10$^{12}$~p$^+$/s. The highest radiation exposure is in the n\_TOF storage area, but still below the applicable limit.}
    \label{7-3_prompt_dose_rate}
\end{figure}
\begin{figure}[h]
    \centering
    \includegraphics[width=\columnwidth]{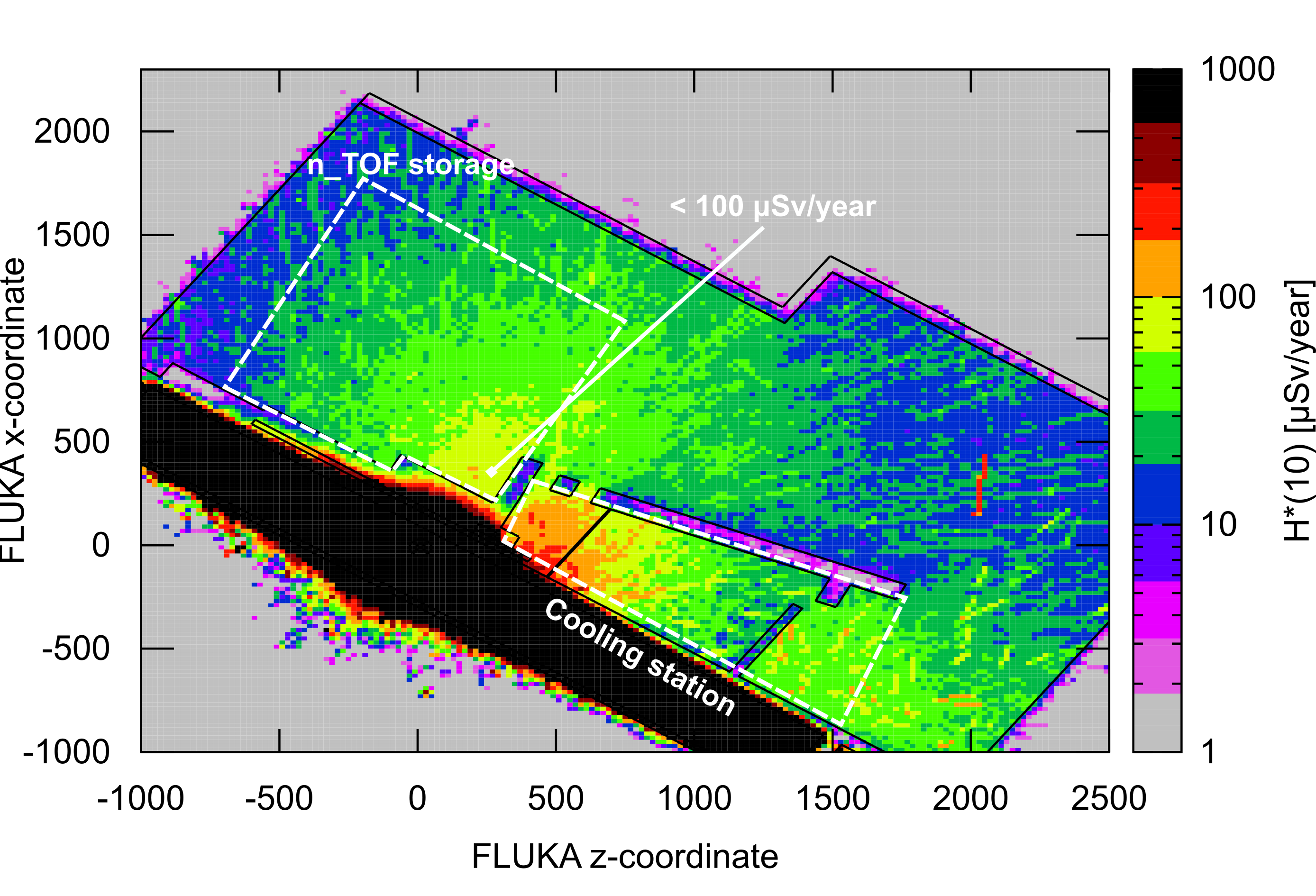}
    \caption{Annual ambient radiation dose equivalent H*(10) in ISR8 for 400 hours of exposure and 2.5$\times$10$^{19}$ protons on target. The annual exposure is always below the limit of 100~\textmu Sv.}
    \label{7-4_yearly_dose}
\end{figure}

\textsc{FLUKA} simulations were performed to assess the ambient dose equivalent, H*(10). Figure~\ref{7-3_prompt_dose_rate} shows the prompt H*(10) rate in the ISR8 area normalized to the average beam intensity of 1.67$\times$10$^{12}$~p$^+$/s: the worst location in terms of radiation exposure is inside the n\_TOF storage area close to the shielding wall. The H*(10) is, however, always lower than 0.3~\textmu Sv/h, well below the applicable limit.
In addition, the simulated H*(10) and the integrated number of protons delivered to the target in a year were used to estimate the exposure of personnel working in ISR8 (400 hours per year considering ISR8 a low-occupancy area). Figure~\ref{7-4_yearly_dose} shows that the annual exposure of personnel in ISR8 is always below the CERN optimization objective of 100~\textmu Sv.
At the time of the construction of the vertical beam line in 2014, the shielding of EAR2 was conservatively designed to take into account a possible increase of the neutron fluence up to a factor of three. Calculations showed that the new neutron fluence is within this threshold.

\subsection{\label{sec:RP:cooling_circuit}Cooling circuit}

From an RP standpoint, one of the critical aspects of the operation of Target~\#2 was the direct contact between the cooling water and the lead core; due to erosion-corrosion and out-diffusion effects, lead spallation products migrated from the target into the cooling water. These radionuclides include not only $\gamma$-emitters like Hg-194, Au-195, and Bi-207, but also $\alpha$-emitters like Gd-148, Po-208, Po-209, and Po-210. The cooling circuit is equipped with cartridges to trap some of these components and reduce the contamination in the circuit. As a consequence, the cartridges represented radiation hot spots and were an important source of external exposure during inspections or maintenance activities in the cooling station; in addition, the presence of contamination in the circuit increased the radiological risk during intervention.

The new nitrogen cooling circuit coupled with HEPA and active carbon filters reduces the presence of lead spallation products and increases the radiological safety during maintenance activities. As described in Sec.~\ref{sec:nitrogen}, the room housing the filters is under dynamic confinement and the extracted air is directed into the target area from where it is released outside, thanks to the preexisting ventilation and radiation monitoring system (see also Sec.~\ref{sec:RP:airact}).
Activation and contamination levels of the nitrogen circuit are monitored by a dedicated beta-radiation detector.

\subsection{\label{sec:RP:airact}Air activation in the target area and environmental releases}

The air surrounding the target area becomes activated due to high-energy hadron interactions. Short-lived positron emitters, like $^{11}$C, $^{13}$N, $^{14,15}$O, are produced in oxygen and nitrogen by spallation reactions. In addition, $^{41}$Ar is generated by thermal neutron capture in the natural argon in air. These radionuclides are the main concern when evaluating the environmental impact of radioactive releases.


The n\_TOF target area is dynamically confined from the adjacent zones; the dynamic confinement is guaranteed by extracting air from the target area at a rate of about 300~m$^3$/h during beam operation, and up to 500~m$^3$/h in case of access. After a filtering stage and radiation monitoring, the air is released into the environment. During the operation of Target~\#2, the effective dose due to environmental releases from the n\_TOF facility was, on average, about 0.6~\textmu Sv per year, well below the CERN optimization objective of 10~\textmu Sv per year.

Following the recent modifications implemented in the target area, i.e. the design and installation of the new target and the new mobile shielding, air activation as well as the impact of environmental releases to the reference population group were assessed and compared with the ones of Target~\#2. The particle spectral fluence for neutrons, charged pions, protons, and photons were simulated by \textsc{FLUKA} simulation package. Then, using \textsc{ActiWiz} (an analytical code developed at CERN~\cite{Vincke14}), the particle spectral fluences were folded with production cross-sections for air to obtain the radionuclide production yields per primary proton impinging on the target. The activity released into the atmosphere all along the irradiation, $A_{rel,i}$, of a given radionuclide $i$ can be calculated using Eq.~\ref{eq:1}~\cite{Agosteo09}:
\begin{equation} \label{eq:1}
A_{rel,i} = Y_i\ I\ \frac{\lambda_i}{\lambda_{i,\mathit{eff}}}\ \frac{Q}{V}\left(t_{\mathit{irr}}-\frac{1-e^{-\lambda_{i,\mathit{eff}}t_{\mathit{irr}}}}{\lambda_{i,\mathit{eff}}}\right)
\end{equation}
where $Y_i$ is the production yield for the radionuclide $i$, $\lambda_i$ its decay constant, $\lambda_{i,\mathit{eff}}$ its effective decay constant (corresponding to the sum of $\lambda_i$ and the air changes provided by the ventilation system per unit time, expressed in the same unit of $\lambda_i$), $I$ the beam intensity, $Q$ the ventilation rate, $V$ the target area volume (about 1200~m$^3$), and $t_{\mathit{irr}}$ the irradiation time (assumed to be 200 days per year).
In case of operation with the new target, the total released activity is about 3.9~TBq per year, which is about 20\% lower than the one of Target~\#2 (i.e. 4.6~TBq per year). In both cases, as shown in Table~\ref{tab:airact}, the released activity is dominated by $^{41}$Ar (45\% for Target~\#3 and 62\% for Target~\#2). For Target~\#3, the production of air spallation products is about 20\% higher; however, the $^{41}$Ar production decreases by more than 70\% because of a lower production of thermal neutrons.

\begin{table}
    \caption{\label{tab:airact} Annual released activity, $A_{rel}$, for the radionuclides mostly contributing to the effective dose for radioactive emission.}
    \begin{tabular}{ r r@{\,}l r@{}l r@{}l l }
    \hline\hline
    \rule{0pt}{3mm}\multirow{2}{*}{\thead{Nuclide}} & \multicolumn{2}{c}{\multirow{2}{*}{\thead{t$_{1/2}$}}} & \multicolumn{4}{c}{$A_{rel}$ (Bq/year)} & \multirow{2}{*}{\thead{$\displaystyle\frac{\textrm{Targ.~\#3}}{\textrm{Targ.~\#2}}$}} \\ [0.3mm] \cline{4-7}
    \rule{0pt}{3mm} & \multicolumn{2}{c}{} & \multicolumn{2}{c}{Targ.~\#2} & \multicolumn{2}{c}{Targ.~\#3} & \\
    \colrule
     \rule{0pt}{3.4mm}$^{11}$C & 20.4 & min & 3.69 & $\times$10$^{11}$ & 4.31 & $\times$10$^{11}$ & 1.17 \\
                      $^{13}$N & 10.0 & min & 9.40 & $\times$10$^{11}$ & 1.15 & $\times$10$^{12}$ & 1.22 \\
                      $^{15}$O &  2.0 & min & 4.13 & $\times$10$^{11}$ & 4.96 & $\times$10$^{11}$ & 1.20 \\
                     $^{41}$Ar &  1.8 & h   & 2.83 & $\times$10$^{12}$ & 1.77 & $\times$10$^{12}$ & 0.63 \\
    \hline\hline
    \end{tabular}
\end{table}

In conclusion, the new target and its shielding reduce the radioactive emissions due to air activation from the n\_TOF facility to the environment.

\subsection{\label{sec:RP:waste}Radioactive waste aspects}

\begin{figure}[h]
    \centering
    \includegraphics[width=\columnwidth]{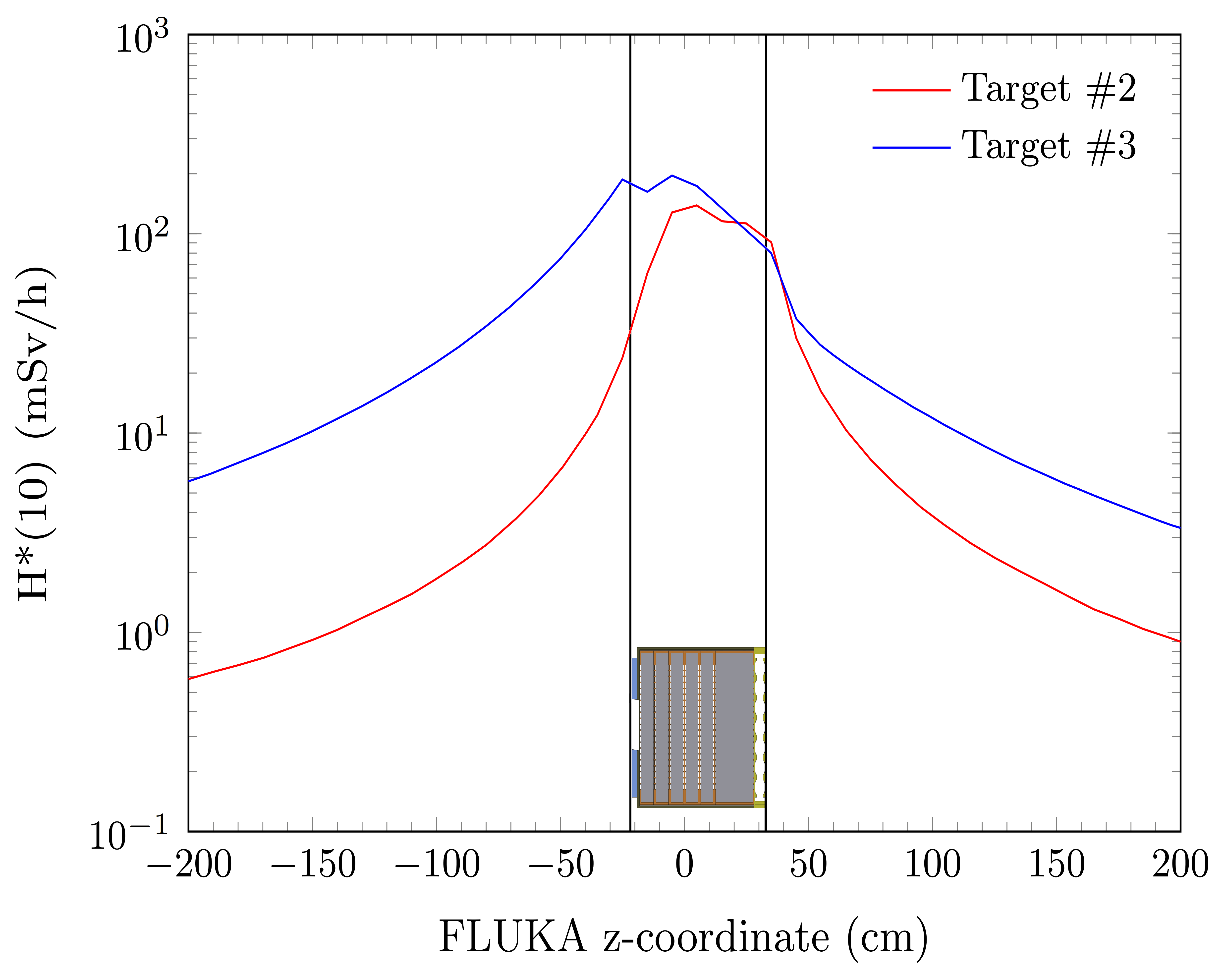}
    \caption{Comparison of the residual dose rate after 10 years of irradiation at 2.5$\times$10$^{19}$ protons on target per year and 3 years of cool-down time for Target~\#2 (red) and Target~\#3 (blue).}
    \label{7-5_residual_dose}
\end{figure}

In view of the final target disposal, the use of aluminum over steel for the target vessel would offer reduced activation and residual dose rate, but lead to additional constraints that would cancel these benefits. Chemical reactions may happen between aluminum and the concrete-based mortar used to prepare the waste, producing hydrogen gas that may damage the package in the long-term~\cite{Pyun00,Mendibide21}. For this reason, it was decided to limit the amount of aluminum in the new target and to build the new vessel in stainless steel. The main cause of the higher residual dose rate of stainless steel is its cobalt content, with the production of $^{60}$Co from thermal neutron capture on $^{59}$Co.
To reduce this contribution, the target vessel is built using stainless steel with cobalt content lower than 0.1\%. \textsc{FLUKA} simulations were performed to compare the residual dose rate after 10 years of irradiation (target lifetime) at 2.5$\times$10$^{19}$ protons on target per year and 3 years of cool-down time (time elapsed between the end of target operation and its shipping to the repository for long term storage). Figure~\ref{7-5_residual_dose} shows that the new target induces a higher maximum residual dose rate by a factor of 10 on the side of the proton beam window of the vessel. These results must be taken into account to plan and optimize the target final disposal.


\section{\label{sec:conclusions}Conclusions}

An upgrade of the n\_TOF facility took place during the CERN Long Shutdown 2 (2019--2021). A major part of the upgrade was the installation of the third-generation neutron spallation target, along with a new nitrogen cooling station and movable target shielding. The new target is based on pure lead cooled by nitrogen gas, a low-cobalt stainless steel vessel, and two water moderators for the two experimental areas of the facility. The nitrogen cooling system replaced the water cooling system of the previous targets, which was the cause of water corrosion and contamination with lead spallation products. This paper presented an overview of the studies leading to the final target design, with details on the physics performance, the mechanical design, the thermo-mechanical aspects, the cooling station, and the radiation protection aspects.

\section*{Acknowledgments}
The authors would like to acknowledge the support and inspiration of the n\_TOF Collaboration in order to reach the current design of this third-generation neutron spallation target. The Project warmly recognizes the financial support provided by the CERN Accelerators Consolidation Project and specifically by M.~Benedikt, R.~Trant and M.~Lamont. R.~Logé wishes to thank PX Group for the generous support to the PX Group Chair. M.~Calviani acknowledges the support and suggestions from the review panel members that helped the project advance and deliver the final product.

\bibliography{references} 

\end{document}